\documentclass[a4paper,10pt]{article}

\pdfoutput=1
\usepackage{jheppub1}
\usepackage{array}
\usepackage{multirow}
\usepackage{amsmath}
\usepackage{makecell}

\usepackage{cancel}
\usepackage{amsmath,amssymb} 
\usepackage{graphicx} 
\usepackage[dvipsnames,x11names]{xcolor} 

\usepackage{hyperref}
\usepackage{cleveref}
\usepackage{float}
\usepackage{subcaption}
\usepackage{comment}

\makeatletter
\newcommand{\footnoteref}[1]{\protected@xdef\@thefnmark{\ref{#1}}\@footnotemark}
\makeatother

\begin{document}

	\preprint{}
	\title{Zeroth law of black hole thermodynamics for higher derivative Proca theories}
	\author[a]{Parthajit Biswas}
    \affiliation[a]{Saha Institute of Nuclear Physics, 1/AF Bidhannagar, Kolkata 700064, India.}

    \author[b]{, Alokananda Kar}
    \author[b]{, and Anowar Shaikh}
    \affiliation[b]{Department of Physics, Indian Institute of Technology (Indian School of Mines) Dhanbad, Jharkhand 826004, India.}
    
	\emailAdd{parthajitbiswas8@gmail.com, 21dr0013@phy.iitism.ac.in, anowar.19dr0016@ap.iitism.ac.in}

%\abstract{We extend the proof of the zeroth law of black hole thermodynamics to arbitrary higher-derivative theories of gravity coupled to Proca fields $-$ {\it i.e.}, vector fields without $U(1)$ gauge invariance $-$ within the framework of effective field theory.} %Our approach treats higher-curvature corrections perturbatively, demonstrating that the surface gravity remains constant across the horizon for stationary black holes, thereby generalizing the zeroth law to this broader theoretical setting.}

\abstract{We prove the constancy of surface gravity across a Killing horizon (not necessarily of bifurcate type) in arbitrary higher curvature theories of gravity coupled to Proca fields $-$ vector fields lacking $U(1)$ gauge invariance $-$ thus generalizing the zeroth law to this broader class of theories. This is achieved within the framework of effective field theory, where higher curvature contributions are treated perturbatively around the leading two derivative theory. The result holds to arbitrary order in the effective field theory expansion. The proof is based on boost-weight arguments; implementing these arguments in the presence of a Proca field introduces subtleties beyond those encountered in the pure-gravity case, which we address here. }

\maketitle

\section{Introduction} \label{sec:intro}

Black holes are fascinating objects, yet in one respect they behave like ordinary matter: the classical properties of their horizons admit a thermodynamic interpretation, where geometric quantities are identified with thermodynamic ones. This correspondence has been rigorously established within Einstein's General Relativity (GR) \cite{Bekenstein:1973ur,Bardeen:1973gs,Hawking:1975vcx}. Since GR is not a UV-complete theory of gravity, it is natural to examine black hole thermodynamics including higher derivative corrections to the Einstein-Hilbert action. These terms generally modify black hole solutions and, consequently, alter the geometric quantities entering thermodynamic relations $-$ thus providing an intriguing setting to test the validity of the laws of black hole thermodynamics beyond GR.

%Beginning with the seminal works of \cite{Iyer:1994ys, Wald:1993nt}, substantial progress has been made \cite{Jacobson:1993vj, Jacobson:1993xs, Sarkar:2013swa, Bhattacharjee:2015yaa} in extending the laws of black hole thermodynamics to higher-curvature theories of gravity.

In \cite{Wald:1993nt,Iyer:1994ys}, the authors demonstrated that {\it the equilibrium state version} of the first law holds universally for arbitrary diffeomorphism invariant higher curvature theories of gravity using the Noether charge method without relying on effective field theory (EFT) approximation. They identified entropy with a geometric quantity which works perfectly well for stationary situations but suffers from well known JKM ambiguities \cite{Jacobson:1993vj} in the dynamical situation. Progress was made in certain restricted cases \cite{Jacobson:1993xs, Sarkar:2013swa, Bhattacharjee:2015yaa}, yet a general proof of the zeroth or second law remained absent. In recent years, significant advances have been made. By treating the dynamics as small perturbations around stationary black hole solutions, the authors of \cite{Wall:2015raa, Bhattacharya:2019qal, Bhattacharyya:2021jhr} have proven the second law for arbitrary higher-curvature theories of gravity, without relying on EFT approximations. These results have since been significantly extended in \cite{Wall:2024lbd, Biswas:2022grc, Bhattacharyya:2022njk, Chandranathan:2022pfx, Hollands:2022fkn, Davies:2022xdq, Davies:2023qaa, Hollands:2024vbe, Visser:2024pwz, Bhattacharyya:2024xpu, Kar:2024dqk}, broadening the scope and applicability of the second law in increasingly general settings. %This was further extended in \cite{Wall:2024lbd, Biswas:2022grc, Bhattacharyya:2022njk, Chandranathan:2022pfx, Hollands:2022fkn, Davies:2022xdq, Davies:2023qaa, Hollands:2024vbe, Visser:2024pwz, Bhattacharyya:2024xpu, Kar:2024dqk}.

This note, on the other hand, focuses on the zeroth law of black hole thermodynamics. {\it The zeroth law of black hole thermodynamics states that the surface gravity and hence the temperature of any stationary black hole is uniform across its event horizon}. This was proven for Einstein's GR in \cite{Bardeen:1973gs}. A fully non perturbative proof of the zeroth law in arbitrary higher curvature theories of gravity remains unavailable. Approaches based on small dynamical perturbations \cite{Wall:2015raa, Bhattacharya:2019qal, Bhattacharyya:2021jhr} are ill suited, since the law pertains exclusively to strictly stationary solutions. Instead, one can invoke an effective field theory (EFT) expansion, treating higher derivative terms as perturbative corrections to the two derivative Einstein-Hilbert action. Within this EFT framework, Bhattacharyya et al. have recently established the zeroth law for general higher curvature gravities in vacuum \cite{Bhattacharyya:2022nqa}, and their proof has been extended to include scalar and $U(1)$ gauge fields in \cite{Davies:2024fut}.

The proofs in \cite{Bhattacharyya:2022nqa, Davies:2024fut} rely heavily on the ``boost-weight" arguments developed in \cite{Wall:2015raa, Bhattacharya:2019qal, Bhattacharyya:2021jhr}. These arguments work nicely for pure gravity and for theories including scalar or $U(1)$ gauge fields $-$ but become a bit subtle when applied to Proca fields (massive vector fields). In this note, we address the subtlety and thereby establish the zeroth law for theories that include Proca fields.

In this note, we aim to establish the zeroth law of black hole thermodynamics for the Einstein-Proca theory, incorporating higher curvature corrections within the framework of EFT. If we consider a {\it gauge field} $B_\mu$ in $D$ dimensions, it has $D-2$ propagating degrees of freedom. Whereas, if we consider a {\it Proca field} $A_\mu$ that does not enjoy $U(1)$ gauge invariance, {\it one more} degree of freedom starts propagating, therefore in $D$ dimension, Proca field $A_\mu$ has $D-1$ propagating degrees of freedom $-$ the time component does not propagate \cite{Heisenberg:2016eld,Heisenberg:2014rta,deRham:2018qqo,Pozsgay:2023wmy,Donoghue:1994dn}. Massive gravity and its ghost-free formulations have been extensively explored as consistent extensions of general relativity~\cite{deRham:2014zqa}.
There has been increasing recent interest in non–gauge vector fields, ranging from the consistency of self-interacting massive vectors~\cite{Banerjee:2025fph} to the cosmological role of Proca fields with non-minimal couplings~\cite{DeFelice:2025ykh, Ozsoy:2023gnl}.

To prove the zeroth law, we work in a coordinate system adapted to the near-horizon geometry. We demonstrate the constancy of surface gravity on a Killing horizon, defined as a codimension-one null hypersurface whose normal vector is Killing, without assuming the horizon to be of bifurcate type. Our strategy involves organizing a specific component of the equations of motion (EOM) order by order in the EFT expansion. We show that the off-shell structure of this component can be expressed in terms of derivatives of the surface gravity with respect to horizon coordinates. Although higher curvature terms can be intricate, we do not require their explicit forms. Instead, by employing ``boost-weight" arguments as developed in \cite{Bhattacharyya:2022nqa}, we abstractly  show this particular structure of the off shell component of the equation of motion within the EFT framework. Finally, imposing the vanishing of the EOM yields the desired result: the surface gravity remains constant across the Killing horizon.

%We prove this by working in a specific choice of the coordinate system adapted to the near-horizon geometry. Then we prove the constancy of surface gravity on a Killing horizon {\it i.e.}, a codimension one null hypersurface whose normal on the hypersurface is a Killing vector, but we do not need to assume that this Killing horizon is of {\it bifurcate type}. We organize a particular component of EOM order by order in EFT parameter and show that their off shell form can be written in a manner directly tied to derivatives of the surface gravity with respect to horizon coordinates. The higher curvature terms can be very complicated but we do not need to use their particular forms rather using ``Boost-Weight" arguments as has been done in \cite{Bhattacharyya:2022nqa}, we can abstractly argue for the zeroth law for arbitrary higher derivative theories within the set up of EFT. Demanding that the EOM vanish then implies the desired constancy of the surface gravity.

We can consider a broader class of vector fields $C_\mu$, in which all $D$ degrees of freedom are dynamical \footnote{Throughout this note, we use the notation $C_\mu$ to denote a vector field whose components each correspond to propagating degrees of freedom. In contrast, $B_\mu$ refers to a standard $U(1)$ gauge field, and $A_\mu$ denotes a Proca field. In $D$ dimensional spacetime, for example, the field $C_\mu$ carries $D$ propagating degrees of freedom, whereas $B_\mu$ and $A_\mu$ possess $D-2$ and $D-1$ propagating degrees of freedom, respectively.}. This generalization allows for additional terms in the Lagrangian. Within this extended framework, we have been able to prove the zeroth law of black hole thermodynamics only for a restricted class of Lagrangians. To avoid confusion and maintain clarity, we refrain from discussing this broader class of theories outside section \S\ref{sec:comments}, where we elaborate on these issues in detail.

%Interestingly, within this extended framework, we have been able to prove the zeroth law of black hole thermodynamics only for a specific combination of leading-order terms. In general, such theories suffer from Ostrogradsky ghost instabilities. However, this particular combination appears to be uniquely ghost-free, ensuring that no propagating Ostrogradsky modes arise. Although we have not established a formal no-go theorem, it seems plausible that the presence of propagating ghosts could invalidate the zeroth law in these more general settings. To avoid confusion and maintain clarity, we refrain from discussing this broader class of theories outside section \S\ref{sec:comments}, where we elaborate on these issues in detail.

The remainder of this paper is organized as follows. In section \S\ref{sec:setup}, we present the effective field theory (EFT) framework, introduce the near-horizon coordinate system, and state our main result. Section \S\ref{sec:proof}, which forms the core of this note, contains the inductive argument establishing the zeroth law to arbitrary perturbative order. In section \S\ref{sec:comments}, we discuss the broader class of vector theories involving $C_\mu$ and offer related observations. Section \S\ref{sec:discussion} summarizes our findings and outlines possible directions for future research. Finally, appendices \ref{app:AA}, \ref{app:boost} and \ref{app:nonProcaL} provide additional computational details and a summary of the “boost-weight” arguments.

\section{Set-up and statement of the problem}\label{sec:setup}
We consider arbitrary higher curvature theories of gravity coupled to a Proca field $A_\mu$. Schematically, we can write our full effective field theory action as follows,
\begin{equation}\label{eq:starting}
S=\int d^Dx\sqrt{-g}\left(\sum_{n=0}^\infty \alpha^{n} \mathcal{L}_{(n)} \right)\,,
\end{equation}
where $\alpha$ is the parameter that controls the effective field theory expansion. The leading order Lagrangian $\mathcal{L}_{(0)}$ has the standard Einstein-Proca theory form
\begin{equation}\label{eq:S0}
\mathcal{S}=\int d^Dx \sqrt{-g}\, \mathcal{L}_{(0)}=\int d^Dx \sqrt{-g}\left(R-\frac{1}{4}F_{\mu\nu}F^{\mu\nu}-\frac{1}{2}m^2 A_\mu A^\mu\right)\,,
\end{equation} 
where,
\begin{equation}
F_{\mu\nu}=\nabla_\mu A_\nu-\nabla_\nu A_\mu\,,
\end{equation}

%\eqref{eq:S0}
%\begin{equation}
%\mathcal{L}_{(0)}=R-\frac{1}{4}F_{\mu\nu}F^{\mu\nu}-\frac{1}{2}m^2 A_\mu A^\mu\,.
%\end{equation}
%In the other case where the temporal component of the vector field is propagating $\mathcal{L}_{(0)}$ has the following form
%\begin{equation}
%\mathcal{L}_{(0)}=R-\frac{1}{4}F_{\mu\nu}F^{\mu\nu}-\frac{1}{2}m^2 A_\mu A^\mu+c_1(\nabla_\mu A_\nu)(\nabla^\nu A^\mu)+c_2(\nabla_\mu A^\mu)^2\,.
%\end{equation}
%Where $c_1$ and $c_2$ are $\mathcal{O}(1)$ constants. 
and, we work in a system of units in which $16\pi G_N$ is set to unity. All the terms in Lagrangian $\mathcal{L}_{(n)}$ should be of the same ``strength'' in effective field theory expansion. In our effective field theory expansion, we organize terms according to their number of derivatives: the more derivatives a term has, the weaker its ``strength". Which implies that in our starting Lagrangian, $m^2 A_\mu A^\mu$ term is of the same order as that of $R$ and $F_{\mu\nu}F^{\mu\nu}$, therefore we will treat $m^2A_\mu A^\mu$ effectively as a two derivative term.

Let's discuss briefly about possible higher order terms. In pure gravity, every term in the action involves an even number of derivatives $-$ for example, the lowest-order contribution is the standard two-derivative Einstein-Hilbert term, with corrections appearing at the four-derivative level, six-derivative level, and so on. However, when a vector field without gauge symmetry is introduced, odd-derivative terms can also arise. In our case $\mathcal{L}_{(1)}$ is a three derivative term which in general will contain the following terms \cite{deRham:2018qqo,deRham:2020yet,Heisenberg:2014rta}
\begin{equation}
\mathcal{L}_{(1)}= a_1 m^2 A^2 (\nabla_\mu A^\mu)+ a_2 \widetilde{F}^{\mu\alpha}{\widetilde{F}^\nu}_{\,\,\,\,\alpha} \nabla_\mu A_\nu+a_3 R (\nabla_\mu A^\mu)+a_4 R_{\mu\nu}\nabla^\mu A^\nu\,.
\end{equation}
where the coefficients $a_i$ are dimensionless constants and $\widetilde{F}_{\mu\nu}$ is dual to Maxwell field strength $F_{\mu\nu}$.

In the next order, the expansion will include terms with up to four derivatives. Specifically, the gravitational sector will include contributions such as
\begin{equation}
R^2, R^{\mu\nu}R_{\mu\nu}, R_{\mu\nu\alpha\beta}R^{\mu\nu\alpha\beta}\,,
\end{equation}
while the vector field sector will include terms like \cite{deRham:2018qqo,deRham:2020yet,Heisenberg:2014rta}
\begin{equation}
m^4 A^4, m^2 A^2 F^{\mu\nu}F_{\mu\nu}, m^2 A^2\left[(\nabla\cdot A)^2-(\nabla_\alpha A_\beta)(\nabla^\beta A^\alpha)\right],\cdots
\end{equation}
along with mixed contributions involving both the curvature tensor $R_{\mu\nu\alpha\beta}$ and the vector field $A_\mu$. We do not aim to provide an exhaustive list here, as the explicit forms of these terms are not essential for the present discussion.

% (see Eqn. (2.32) of \cite{Pozsgay:2023wmy} for the corresponding pure vector field terms), along with mixed contributions involving both the curvature tensor $R_{\mu\nu\alpha\beta}$ and the vector field $A_\mu$. We do not present their explicit forms here, as they are not essential for the current discussion.
%%%%
%For example, our $\mathcal{L}_{(1)}$ which is a four derivative term contains the following terms
%\begin{equation}\label{eq:L1ex}
%\mathcal{L}_{(1)}=c_1^{(1)} R^2+c^{(1)}_2 R_{\mu\nu}R^{\mu\nu}+c^{(1)}_3 R_{\mu\nu\rho\sigma}R^{\mu\nu\rho\sigma}+c^{(1)}_4 R A_\mu A^\mu+c^{(1)}_5 R_{\mu\nu}A^\mu A^\nu+c^{(1)}_6 R F_{\mu\nu}F^{\mu\nu}+\cdots
%\end{equation}
%where, $c_1^{(1)}, c_2^{(1)}, \cdots $ are $\mathcal{O}(1)$ constants.\\\\
%%

If we compute the equations of motion from \eqref{eq:starting}, they will take the form of a perturbative expansion in powers of $\alpha$. Let us schematically write the equations of motion for the metric and the vector field as follows:
\begin{equation}\label{eq:EoM}
\begin{split}
E_{\mu\nu}&=E^{(0)}_{\mu\nu}+\alpha\,E^{(1)}_{\mu\nu}+\alpha^2 E^{(2)}_{\mu\nu}+\cdots\\
E_{\mu}&=E^{(0)}_{\mu}+\alpha\,E^{(1)}_{\mu}+\alpha^2 E^{(2)}_{\mu}+\cdots\,,
\end{split}
\end{equation}
where $E^{(0)}_{\mu\nu}$ and $E^{(0)}_\mu$ are the part of the equations of motion coming from $\mathcal{L}_{(0)}$ and similarly for all the other $E^{(n)}_{\mu\nu}$ and $E^{(n)}_\mu$.

As mentioned earlier in the introduction, the zeroth law of black hole thermodynamics pertains to stationary solutions. Suppose, $g_{\mu\nu}^{(bh)}$ and $A_{\mu}$ is a stationary solution of $E_{\mu\nu}=0$ and $E_{\mu}=0$. Therefore, $g_{\mu\nu}^{(bh)}$ and $A_\mu$ will also admit expansion in terms of effective field theory parameter $\alpha$ 
\begin{equation}\label{eq:bhmetric}
\begin{split}
g_{\mu\nu}^{(bh)}&=g_{\mu\nu}^{(0)}+\alpha\,g_{\mu\nu}^{(1)}+\cdots\,,\\
A_\mu&=A^{(0)}_\mu+\alpha\, A^{(1)}_\mu+\cdots\,.
\end{split}
\end{equation} 
As the metric $g_{\mu\nu}^{(bh)}$ is stationary, it admits an asymptotically timelike Killing vector field, which we will denote by $\xi^\mu$. Let us further assume that the event horizon $\Sigma$ of $g_{\mu\nu}^{(bh)}$ is Killing (but not assumed to be of {\it bifurcate} type) for the Killing vector field  $\xi^\mu$. As discussed in \cite{Bhattacharyya:2021jhr,Bhattacharyya:2022nqa}, it is always possible to choose a coordinate system in which the metric near the Killing horizon $\Sigma$ takes the following form:
\begin{equation}\label{eq:metric}
\begin{split}
ds^2 &= g^{(bh)}_{\mu\nu} dX^\mu dX^\nu \\ &= 2\, d\tau\, d\rho - \rho\, X(\rho, x^i)\, d\tau^2 + 2\rho\, \omega_i(\rho, x^i)\, d\tau\, dx^i + h{ij}(\rho, x^i)\, dx^i dx^j\,.
\end{split}
\end{equation}
Here, $\xi^\mu = \partial_\tau$ is the Killing vector, implying that none of the metric components depend on $\tau$. In the coordinate system $(\tau, \rho, x^i)$, the Killing horizon $\Sigma$ is located at $\rho = 0$, where the norm of the Killing vector vanishes.
The vector potential $A_\mu$ must also satisfy the stationarity condition $\mathcal{L}_\xi A_\mu = 0$, which leads to \begin{equation}\label{eq:Astat}
\begin{split}
\mathcal{L}_\xi A_\mu = 0 \quad \Rightarrow \quad \partial_\tau A_\mu = 0\,.
\end{split}
\end{equation}
Thus, the components of $A_\mu$ are independent of $\tau$, and can be expressed as functions of $\rho$ and $x^i$
\begin{equation}
A_\mu = A_\mu(\rho, x^i)\,.
\end{equation}
To prove that the surface gravity is constant over the Killing horizon, we start with its definition:
\begin{equation}
\kappa = \sqrt{-\frac{1}{2} (\nabla_\mu \xi_\nu)(\nabla^\mu \xi^\nu)|_{\rho=0}} \,.
\end{equation}
By evaluating $\kappa$ for the metric \eqref{eq:metric}, it is found to be
\begin{equation}
\kappa = \frac{1}{2} X(\rho=0, x^i) \,.
\end{equation}
To establish the constancy of the surface gravity over the horizon in our coordinate system, we need to show that
\begin{equation}
\partial_\tau \kappa = 0 \,, \quad \text{and} \quad \partial_i \kappa = 0 \,.
\end{equation}
Since $\partial_\tau$ is a Killing vector, none of the metric components depend on $\tau$. Consequently, $X$ has no dependence on $\tau$, and hence $\partial_\tau \kappa = 0$ is satisfied trivially. To complete the proof of the zeroth law, it remains to show that $\partial_i\kappa=0$, in our coordinate system we have to show that
\begin{equation}\label{eq:goal}
\partial_i X\Big|_{\rho=0}=0\,.
\end{equation}
We will demonstrate this by utilizing two specific components $E_{\tau i}$ and $E_{\tau}$ of the equations of motion, analyzed order by order within the framework of the effective field theory expansion.

\section{Proof of the zeroth law}\label{sec:proof}
As we have discussed in section \ref{sec:setup}, our goal is to prove eqn. \eqref{eq:goal} using equation of motion within the framework of effective field theory. We write our equation of motion in effective field theory expansion as written in eqn. \eqref{eq:EoM}
\begin{equation}
\begin{split}
E_{\mu\nu}&=E^{(0)}_{\mu\nu}+\alpha\,E^{(1)}_{\mu\nu}+\alpha^2 E^{(2)}_{\mu\nu}+\cdots\,,\\
E_{\mu}&=E^{(0)}_{\mu}+\alpha\,E^{(1)}_{\mu}+\alpha^2 E^{(2)}_{\mu}+\cdots\,.
\end{split}
\end{equation}
The solution of $E_{\mu\nu}=0$ and $E_\mu=0$ which we denote by $g_{\mu\nu}^{(bh)}$ and $A_\mu$ can also be written in effective field theory expansion as we have written in \eqref{eq:bhmetric}
\begin{equation}\label{eq:met_exp}
\begin{split}
g_{\mu\nu}^{(bh)}&=g_{\mu\nu}^{(0)}+\alpha\,g_{\mu\nu}^{(1)}+\cdots\,,\\
A_\mu&=A^{(0)}_\mu+\alpha\, A^{(1)}_\mu+\cdots\,.
\end{split}
\end{equation} 
If we write our metric in the horizon adapted coordinate system as written in \eqref{eq:metric} then eqn. \eqref{eq:met_exp} implies that $X$, $\omega_i$, $h_{ij}$ and also $A_\mu$ will admit expansion in terms of $\alpha$
\begin{equation}\label{eq:gmunuexpansion}
\begin{split}
X(\rho,x^i)&=X^{(0)}(\rho,x^i)+\alpha\,X^{(1)}(\rho,x^i)+\alpha^2\,X^{(2)}(\rho,x^i)+\cdots\,,\\
\omega_i(\rho,x^i)&=\omega_i^{(0)}(\rho,x^i)+\alpha\,\omega_i^{(1)}(\rho,x^i)+\alpha^2\,\omega_i^{(2)}(\rho,x^i)+\cdots\,,\\
h_{ij}(\rho,x^i)&=h_{ij}^{(0)}(\rho,x^i)+\alpha\,h_{ij}^{(1)}(\rho,x^i)+\alpha^2\,h_{ij}^{(2)}(\rho,x^i)+\cdots\,,\\
A_\mu(\rho,x^i)&=A_\mu^{(0)}(\rho,x^i)+\alpha\,A_\mu^{(1)}(\rho,x^i)+\alpha^2\,A_\mu^{(2)}(\rho,x^i)+\cdots\,.
\end{split}
\end{equation}
At the leading order, \textit{i.e.}, at $\mathcal{O}(\alpha)^0$, the metric $g^{(0)}_{\mu\nu}$ is an exact solution to $E^{(0)}_{\mu\nu} = 0$. Assuming that the equation of motion at the leading order, $E^{(0)}_{\mu\nu} = 0$, has been solved, it follows that $g^{(0)}_{\mu\nu}$ is known exactly. At the first subleading order, \textit{i.e.,} at $\mathcal{O}(\alpha)^1$, the equation of motion is
\begin{equation}
E^{(0)}_{\mu\nu}+\alpha\, E^{(1)}_{\mu\nu}=\mathcal{O}(\alpha)^2\,.
\end{equation}
At this order the metric also has the following form
\begin{equation}
g_{\mu\nu}^{(bh)}=g_{\mu\nu}^{(0)}+\alpha\, g_{\mu\nu}^{(1)}+\mathcal{O}(\alpha)^2\,.
\end{equation}
Here $g_{\mu\nu}^{(0)}$ are some known functions, therefore the variables of the equation of motion at the first subleading order are $g^{(1)}_{\mu\nu}$. Equations we have to solve to determine $g^{(1)}_{\mu\nu}$ are the followings
\begin{equation}\label{eq:1stsub}
\Big[E^{(0)}_{\mu\nu}+\alpha\, E^{(1)}_{\mu\nu}\Big]\Big[g^{(0)}_{\mu\nu}+\alpha g^{(1)}_{\mu\nu}\Big]=\mathcal{O}(\alpha)^2\,.
\end{equation}
All the terms in the above equation are $\mathcal{O}(\alpha)$ terms as we have solved the equation of motion at the leading $\mathcal{O}(\alpha)^0$ order. The above equation can be written as
\begin{equation}\label{eq:1st_decom}
E^{(0)}_{\mu\nu}\Big[g^{(0)}_{\mu\nu}+\alpha\, g^{(1)}_{\mu\nu}\Big]+\alpha\, E^{(1)}_{\mu\nu}\Big[g^{(0)}_{\mu\nu}\Big]=\mathcal{O}(\alpha)^2\,.
\end{equation}
%The first part of the above equation is the linearization of $E^{(0)}_{\mu\nu}$ around $g^{(0)}_{\mu\nu}$ assuming $g^{(1)}_{\mu\nu}$ a small perturbations, this we will call \textit{homogeneous part} of the above differential equation. When we will try to solve the equation of motion at arbitrary order this part of the differential equation will be universal as we will explain below. $E^{(1)}_{\mu\nu}$ depends on higher curvature terms $\mathcal{L}_{(1)}$ in the Lagrangian, once $\mathcal{L}_{(1)}$ is specified  $E^{(1)}_{\mu\nu}$ is some known covariant tensor, we have to compute this on the zeroth order solution $g^{(0)}_{\mu\nu}$, this part of the differential equation we will call \textit{source part}. The source part of the differential equation is not universal it depends on the details of the theory and the particular order that we are considering. As we will discuss below, we really do not need to compute source part of the differential equation, this we will be able to handle using some abstract arguments this was the main point of the work \cite{Bhattacharyya:2022nqa}.

The first part in the above equation represents the linearization of $E^{(0)}_{\mu\nu}$ around $g^{(0)}_{\mu\nu}$, treating $g^{(1)}_{\mu\nu}$ as a small perturbation. We will refer to this as the \textit{homogeneous part} of the differential equation. As we will explain below, this part of the equation is universal when solving the equations of motion at any arbitrary higher order. However, the homogeneous part depends on the specific form of $\mathcal{L}_{(0)}$. It is universal with respect to effective field theory perturbations, meaning that once $\mathcal{L}_{(0)}$ is specified, the form of the homogeneous part will remain the same across all higher orders.

The term $E^{(1)}_{\mu\nu}$, on the other hand, depends on the higher curvature contributions $\mathcal{L}_{(1)}$ in the Lagrangian. Once $\mathcal{L}_{(1)}$ is specified, $E^{(1)}_{\mu\nu}$ becomes a known covariant tensor, which we have to evaluate on the zeroth-order solution $g^{(0)}_{\mu\nu}$. This part of the equation will be referred to as the \textit{source part} which acts like source in the inhomogeneous partial differential equations for $g^{(1)}_{\mu\nu}$.  At any arbitrary higher order, the equation of motion can always be decomposed into two parts: a homogeneous part and a source part. Unlike the homogeneous part, the source part is not universal $-$ it varies depending on the specific theory and the order being considered. As we will demonstrate, there is no need to compute the source term directly: it can instead be addressed using abstract arguments, as developed in \cite{Bhattacharyya:2022nqa}.

The proof can be systematically extended to arbitrary higher order using {\it the method of induction}. According to this method, if the proof holds at the leading order, then it will continue to hold at arbitrary higher orders $-$ provided we can demonstrate its validity at the $(n+1)$th order assuming it holds at the $n$th order. Notably, this method does not require us to explicitly verify the proof at the first subleading order. However, for illustrative purposes, we will present the proof at the first subleading order, which captures all essential components of our argument in a simpler setting.

Let us assume that the equations of motion for both the metric and the vector field have been solved up to order $\mathcal{O}(\alpha)^n$. At the next order, $\mathcal{O}(\alpha)^{n+1}$, the unknown functions are $g_{\mu\nu}^{(n+1)}$ and $A_\mu^{(n+1)}$. At this order, the equations of motion can be systematically decomposed into two parts $-$ one involving the unknown functions $g_{\mu\nu}^{(n+1)}$ and $A_\mu^{(n+1)}$, which possesses a universal structure across all orders and is referred to as the {\it homogeneous part}, and another that depends on the specific form of the higher-curvature gravity theory, rendering it non-universal and termed as the {\it source part}. It is important to note that the source part is independent of the unknown functions $g_{\mu\nu}^{(n+1)}$ and $A_\mu^{(n+1)}$.\\\\
Homogeneous part at the order $\mathcal{O}(\alpha)^{n+1}$ has the following structure
\begin{equation}
E_{\mu\nu}^{(0)}\left[g_{\mu\nu}^{(0)}+\alpha^{n+1} g_{\mu\nu}^{(n+1)}, A_\mu^{(0)}+\alpha^{n+1}A_\mu^{(n+1)}\right]\,,
\end{equation}
which is essentially the linearization of $E_{\mu\nu}^{(0)}$ around $(g_{\mu\nu}^{(0)},A_\mu^{(0)})$ assuming $\alpha^{n+1} g_{\mu\nu}^{(n+1)}$ and $\alpha^{n+1}A_\mu^{(n+1)}$ as linearized fluctuations.\\\\
On the other hand, the source part has the following structure
\begin{equation}
\left[E_{\mu\nu}^{(0)}+\cdots+\alpha^{n+1}E_{\mu\nu}^{(n+1)}\right]\left[g_{\mu\nu}^{(0)}+\cdots+\alpha^{n} g_{\mu\nu}^{(n)}, A_\mu^{(0)}+\cdots+\alpha^{n}A_\mu^{(n)}\right]\,.
\end{equation}
At the current order $\mathcal{O}(\alpha)^{n+1}$, all metric components up to $g_{\mu\nu}^{(n)}$ and vector field components up to $A_{\mu}^{(n)}$ are known functions. Consequently, once the higher-curvature theory is specified, the source part is, in principle, determined. However, explicitly computing these source terms can be exceedingly tedious. Following \cite{Bhattacharyya:2022nqa}, we will show that such a computation is unnecessary; instead, the source part can be treated abstractly using boost weight arguments.

%\subsection{Case I: Temporal component of $A_\mu$ is non-dynamical}\label{subsec:caseI}
%In this subsection, we will consider vector field $A_\mu$ which \textit{does not have dynamical temporal component}. This is referred to as Proca theory in the literature. We can write our full effective field theory action as follows
%\begin{equation}
%S=\int d^Dx\sqrt{-g}\left( \mathcal{L}_{(0)}+\alpha\,\mathcal{L}_{(1)}+\cdots\right)\,.
%\end{equation}
%As the temporal component of $A_\mu$ does not propagate, the kinetic term for $A_\mu$ at the leading order can only have $F_{\mu\nu}F^{\mu\nu}$ as we have discussed in the introduction
%\begin{equation}\label{eq:Ept}
%\mathcal{L}_{(0)}=R-\frac{1}{4}F_{\mu\nu}F^{\mu\nu}-\frac{1}{2}m^2 A_\mu A^\mu\,.
%\end{equation}
%%$\mathcal{L}_{(1)}$, for example, can be of the following form
%%\begin{equation}
%%\mathcal{L}_{(1)}=c_1 R^2+c_2 R_{\mu\nu}R^{\mu\nu}+c_3 R_{\mu\nu\rho\sigma}R^{\mu\nu\rho\sigma}+c_4 R A_\mu A^\mu+c_5 R_{\mu\nu}A^\mu A^\nu+c_6 R F_{\mu\nu}F^{\mu\nu}+\cdots
%%\end{equation}
%In this case, none of the $\mathcal{L}_{(n)}$ should include any terms having derivative acting on $A_\mu$ other than through $F_{\mu\nu}$.

\subsection{Proof at the leading order}\label{subsec:leading}
Now, we will prove constancy of surface gravity at the leading order that is for Einstein-Proca theory defined by the Lagrangian \eqref{eq:S0}. If we compute equations of motion form \eqref{eq:S0} we get
\begin{equation}\label{eq:eoms}
\begin{split}
E_{\mu\nu}^{(0)}&\equiv R_{\mu\nu}-\frac{1}{2}g_{\mu\nu}R+\frac{1}{8}g_{\mu\nu} F_{\alpha \beta} F^{\alpha \beta}-\frac{1}{2}g^{\alpha\beta}F_{\mu\alpha}F_{\nu\beta}+\frac{1}{4}m^2g_{\mu\nu} A^\alpha A_\alpha-\frac{1}{2} m^2 A_{\mu}A_{\nu}=0\,,\\
E^{(0)}_\mu&\equiv\nabla^\nu F_{\nu\mu}-m^2 A_\mu=0\,.
\end{split}
\end{equation}
Following \cite{Bhattacharyya:2022nqa}, we now consider the $\tau i$ component of the eqution of motion. In our horizon adapted coordinate system the zeroth order metric can be written as
\begin{equation}\label{eq:met0}
\begin{split}
ds^2&=2 d\tau~ d\rho-\rho X^{(0)}(\rho,x^i) d\tau^2+2 \rho~ \omega_i^{(0)}(\rho,x^i) d\tau d x^i+h_{ij}^{(0)}(\rho,x^i) dx^i dx^j\,.
\end{split}
\end{equation}
Now, we would compute $\tau i$ component of $E_{\mu\nu}^{(0)}$ on the horizon $\rho=0$
\begin{equation}
\begin{split}
E_{\tau i}^{(0)}\left[g^{(0)}_{\mu\nu}, A^{(0)}_\mu\right]_{\rho=0}&=\left(R^{(0)}_{\tau i}-\frac{1}{2}g^{\alpha\beta}F^{(0)}_{\tau \alpha}F^{(0)}_{i\beta}-\frac{1}{2} m^2 A^{(0)}_{\tau}A^{(0)}_{i}\right)_{\rho=0}\,.
\end{split}
\end{equation}
We can compute $R_{\tau i}\Big|_{\rho=0}$ on the metric \eqref{eq:met0}. The answer is
\begin{equation}\label{eq:Rti0}
R_{\tau i}\left[g_{\mu\nu}^{(0)}\right]\Big|_{\rho=0}=-\frac{1}{2}\partial_i X^{(0)}\,.
\end{equation}
See appendix \ref{app:details1} for details. Using \eqref{eq:Rti0} and the stationarity condition \eqref{eq:Astat} we get
\begin{equation}
\begin{split}
E_{\tau i}^{(0)}\left[g^{(0)}_{\mu\nu}, A^{(0)}_\mu\right]_{\rho=0}=-\frac{1}{2}\left[\partial_i X^{(0)}-g^{\alpha\beta}\left(\partial_\alpha A_\tau^{(0)}\right)F_{i\beta}^{(0)}+m^2 A^{(0)}_{\tau}A_{i}^{(0)}\right]_{\rho=0}\,.
\end{split}
\end{equation}
Using the expressions of inverse metric (appendix \ref{app:details1}), we get
\begin{equation}\label{eq:Eti02}
\begin{split}
E_{\tau i}^{(0)}\left[g^{(0)}_{\mu\nu}, A^{(0)}_\mu\right]_{\rho=0}=-\frac{1}{2}\left[\partial_i X^{(0)}-\left(\partial_\rho A_\tau^{(0)}\right)\left(\partial_i A_\tau^{(0)}\right)-h^{jk}\left(\partial_j A_\tau^{(0)}\right)F_{ik}^{(0)}+m^2 A^{(0)}_{\tau}A_{i}^{(0)}\right]_{\rho=0}\,.
\end{split}
\end{equation}
%\newpage
%To prove the zeroth law, now we will use $E_{\tau\tau}^{(0)}=0$. At the leading order it is given by
%\begin{equation}
%E_{\tau\tau}^{(0)}\left[g^{(0)}_{\alpha\beta},A^{(0)}_\alpha\right]=-\frac{1}{2}\Big[ h^{ij}_{(0)}F_{\tau i}^{(0)}F_{\tau j}^{(0)}+ m_{(0)}^2 \left(A_\tau^{(0)}\right)^2\Big]_{\rho=0}=0
%\end{equation}
%See appendix \ref{app:details1} for the derivation. Both terms on the right-hand side are manifestly positive-definite, as they are squared quantities. Therefore, each must vanish independently, leading to the condition
%\begin{equation}\label{eq:imp1}
%A_\tau^{(0)}\Big|_{\rho=0}=0\,.
%\end{equation}
%%
%%%%
%\newpage
To establish the zeroth law, $\tau i$ component of the metric equation of motion alone is not enough, it turns out that we also have to consider the $\tau$ component of the vector equation of motion (second equation in \eqref{eq:eoms})
%\begin{equation}
%E^{(0)}_\tau\left[g^{(0)}_{\mu\nu}, A^{(0)}_\mu\right]_{\rho=0}=\left(\nabla^\nu_{(0)} F^{(0)}_{\nu\tau}-m^2 A^{(0)}_\tau\right)_{\rho=0}\,,
%\end{equation} 
%where $\nabla^\nu_{(0)}$ is the covariant derivative with respect to $g^{(0)}_{\mu\nu}$. Now, if we evaluate the right hand side on the metric \eqref{eq:met0}, we will get
\begin{equation}\label{eq:int1}
E^{(0)}_\tau\left[g^{(0)}_{\mu\nu}, A^{(0)}_\mu\right]_{\rho=0}=\left(h^{ij}_{(0)}\mathcal{D}^{(0)}_i F^{(0)}_{j\tau}-m^2 A^{(0)}_\tau\right)_{\rho=0}\,,
\end{equation}
where $\mathcal{D}_i^{(0)}$ is the covariant derivative with respect to the metric $h_{ij}^{(0)}$ (see appendix \ref{app:Etau} for the derivation of the above equation). Using $\partial_\tau A_j=0$ we can write
\begin{equation}\label{eq:step1}
\begin{split}
F_{j\tau}^{(0)}&=\partial_j A_\tau^{(0)}\\
&=\mathcal{D}^{(0)}_j A_\tau^{(0)}\,.
\end{split}
\end{equation}
We now multiply \eqref{eq:int1} with $\sqrt{h^{(0)}}A_\tau^{(0)}$ and integrate over a $D-2$ dimensional compact spatial manifold $C$
\begin{equation}\label{eq:step2}
\begin{split}
&\int_C d^{D-2}x \sqrt{h^{(0)}} A_\tau^{(0)}\left[h^{ij}_{(0)}\mathcal{D}^{(0)}_i \mathcal{D}^{(0)}_j A_\tau^{(0)}-m^2 A^{(0)}_\tau\right]_{\rho=0}=0\\
\Rightarrow&\int_C d^{D-2}x \sqrt{h^{(0)}} \left[-h^{ij}_{(0)}\left(\mathcal{D}^{(0)}_i A_\tau^{(0)}\right)\left(\mathcal{D}^{(0)}_j A_\tau^{(0)}\right)-m^2 \left(A^{(0)}_\tau\right)^2\right]_{\rho=0}=0\\
\end{split}
\end{equation}
where we have used the fact that the total derivative term vanishes as we are integrating on a compact manifold. Now, both the terms are full squared term, therefore both of them has to vanish separately. Which gives $A_\tau^{(0)}\Big|_{\rho=0}=0$ which in turn implies $A_\tau^{(0)}$ can be written as
\begin{equation}\label{eq:imp1}
A_\tau^{(0)}=\rho\, \psi_\tau^{(0)}(\rho,x^i)\,.
\end{equation}
Similar computations have been used for the proof of the zeroth law in the context of gauge fields \cite{Davies:2024fut}.
%Using this we see that the first term $h^{ij}_{(0)}\left(\mathcal{D}^{(0)}_i A_\tau^{(0)}\right)\left(\mathcal{D}^{(0)}_j A_\tau^{(0)}\right)$ also vanishes. Though Iain Davies has not considered Proca theory he has used similar computations for gauge field equation of motion \cite{Davies:2024fut}. We could have derived the result \eqref{eq:imp1} using $\tau\tau$ componnet of $E_{\mu\nu}^{(0)}$ (see appendix \ref{app:Ett}). But here we are using $E_\tau$ component to derive $A_\tau^{(0)}=0$ because this can be generalized to higher order to show $A^{(k)}_\tau=0$ as we will discuss in the next subsection.
Substituting eqn. \eqref{eq:imp1} into \eqref{eq:Eti02}, we obtain
\begin{equation}\label{eq:dix0}
\partial_i X^{(0)}\Big|_{\rho=0}=0\,,
\end{equation}
which completes our proof of the zeroth law at the leading order. 
\subsection{Proof at the first subleading order}\label{subsec:subleading}
As we have discussed before, at any particular order, we will always divide the equation of motion into two parts : an {\it universal homogeneous part} and a {\it theory dependent source part}.
Let us first compute the homogeneous part. From eqn. \eqref{eq:1st_decom}, we know that the homogeneous part is given by $E^{(0)}_{\mu\nu}\Big[g^{(0)}_{\mu\nu}+\alpha\, g^{(1)}_{\mu\nu}\Big]$. To evaluate this, we will assume that $g^{(0)}_{\mu\nu}$ satisfies the equation of motion $E^{(0)}_{\mu\nu}=0$.
In our horizon adapted coordinate system the metric up to first subleading order can be written as
\begin{equation}
\begin{split}
ds^2&=2 d\tau~ d\rho-\rho \left(X^{(0)}(\rho,x^i)+\alpha\, X^{(1)}(\rho,x^i)\right) d\tau^2+2 \rho\left(\omega_i^{(0)}(\rho,x^i)+\alpha\,\omega_i^{(1)}(\rho,x^i)\right) d\tau d x^i\\
&+\left(h_{ij}^{(0)}(\rho,x^i)+\alpha\,h_{ij}^{(1)}(\rho,x^i)\right) dx^i dx^j\,.
\end{split}
\end{equation}
Now, we would compute $\tau i$ component of $E_{\mu\nu}^{(0)}$ on the horizon $\rho=0$
\begin{equation}\label{eq:E0taui}
\begin{split}
&E_{\tau i}^{(0)}\left[g^{(0)}_{\mu\nu}+\alpha\,g^{(1)}_{\mu\nu}, A^{(0)}_\mu+\alpha\,A_\mu^{(1)}\right]_{\rho=0}\\
&=\alpha\,\partial_i X^{(1)}-\frac{1}{2}\alpha\, g^{\alpha\beta}F^{(0)}_{\tau \alpha}F_{i\beta}^{(1)}-\frac{1}{2}\alpha\,g^{\alpha\beta}F^{(1)}_{\tau \alpha}F_{i\beta}^{(0)}-\frac{1}{2}\alpha\, m^2 A^{(0)}_{\tau}A^{(1)}_{i}-\frac{1}{2}\alpha\, m^2 A^{(1)}_{\tau}A^{(0)}_{i}+\mathcal{O}(\alpha)^2\\
&=\alpha\,\partial_i X^{(1)}+\frac{1}{2}\alpha\left(\partial_\rho A_\tau^{(0)}\right)\left(\partial_i A_\tau^{(1)}\right)+\frac{1}{2}\alpha\, h^{jk}\left(\partial_j A_\tau^{(1)}\right)F_{ik}^{(0)}-\frac{1}{2}\alpha\, m^2 A^{(1)}_{\tau}A^{(0)}_{i}+\mathcal{O}(\alpha)^2\,,\\
\end{split}
\end{equation}
where in the second line we have used the expression of $R_{\tau i}$ form appendix \ref{app:details1}. In the third line, we have used eqn. \eqref{eq:Astat} and eqn. \eqref{eq:imp1}. This will be the expression of the homogeneous part at any arbitrary $k$-th order, we just have to replace the $A_\tau^{(1)}$ with $A_\tau^{(k)}$, that is the reason we called the homogeneous part universal.

Now, we turn to the computation of the source term. As previously noted, its structure depends on the details of the underlying theory. However, by employing abstract arguments based on boost-weight, we will demonstrate that the source term vanishes. In this subsection, we will illustrate this result at the first subleading order. In the following subsection, we will extend the proof to arbitrary higher orders.

Although it is possible to directly present the general analysis for arbitrary higher orders without discussing the first subleading case, we choose to begin with the first subleading order for clarity and explicitness. For this discussion, we will closely follow the approach of \cite{Bhattacharyya:2022nqa}. Now, from \eqref{eq:dix0} we can write $X^{(0)}$ as
\begin{equation}
X^{(0)}(\rho,x^i)=C^{(0)}+\rho\,F^{(0)}(\rho,x^i)\,,
\end{equation}
where, $C^{(0)}$ is some constant. Therefore the zeroth order metric becomes
\begin{equation}
\begin{split}
ds^2&=2 d\tau~ d\rho-\rho \left(C^{(0)}+\rho\,F^{(0)}(\rho,x^i)\right) d\tau^2+2 \rho~ \omega_i^{(0)}(\rho,x^i) d\tau d x^i+h_{ij}^{(0)}(\rho,x^i) dx^i dx^j\,.
\end{split}
\end{equation}
Now following \cite{Bhattacharyya:2022nqa,Bhattacharya:2019qal}, we will do the following coordinate transformation from $x^\mu=\{\tau,\rho,x^i\}$ to $\tilde{x}^\mu=\{v,r,x^i\}$
\begin{equation}\label{eq:coord_trans}
\begin{split}
v=\frac{2}{C^{(0)}}\exp\left({\frac{C^{(0)}}{2}\tau}\right)\,, \quad r=\rho\,\exp\left(-{\frac{C^{(0)}}{2}\tau}\right)\,,
\end{split}
\end{equation}
where, $v$ is the affine parameter along the null generator of the Killing horizon. The inverse transformation is the following
\begin{equation}
\tau=\frac{2}{C^{(0)}}\log \left(\frac{C^{(0)}}{2}v\right),\quad \rho=\frac{C^{(0)}}{2}r\,v\,.
\end{equation}
The infinitesimal distance in this new coordinate take the following form
\begin{equation}\label{eq:0thaff}
ds^2= 2dv\,dr-r^2 F^{(0)}\left(C^{(0)}rv/2\right)dv^2+r\,\omega_i^{(0)}\left(C^{(0)}rv/2\right)dv\, dx^i+h_{ij}^{(0)}\left(C^{(0)}rv/2\right) dx^i dx^j\,.
\end{equation}
The vector field $A_\mu$ also transforms under this coordinate change according to
\begin{equation}
\tilde{A}^{(0)}_\mu(\tilde{x}^\alpha)=\frac{\partial x^\nu}{\partial \tilde{x}^\mu} A^{(0)}_\nu(x^\alpha)\,.
\end{equation}
Component-wise this yields
\begin{equation}\label{eq:Vmup}
\begin{split}
\tilde{A}_v^{(0)}&=\frac{2}{C^{(0)}v}A_\tau^{(0)}\left(C^{(0)} rv/2,x^i\right)+\frac{C^{(0)}}{2}r A_\rho^{(0)}\left(C^{(0)} rv/2,x^i\right)\,,\\
\tilde{A}_r^{(0)}&=\frac{C^{(0)}}{2}v\, A^{(0)}_\rho\left(C^{(0)} rv/2,x^i\right)\,,\\
\tilde{A}_i^{(0)}&=A_i^{(0)}\left(C^{(0)} rv/2,x^i\right)\,.
\end{split}
\end{equation}
From \eqref{eq:1st_decom}, we know that to compute the source part at first subleading order, we have to compute $\alpha\, E^{(1)}_{\mu\nu}\Big[g^{(0)}_{\mu\nu}\Big]$. We have brought the zeroth order metric in the particular form \eqref{eq:0thaff}. Now, $E^{(1)}_{\mu\nu}$ is just a covariant tensor constructed out of Riemann tensor and their derivatives and contractions. It will transform according the way a two index tensor would transform. Therefore, we have
\begin{equation}\label{eq:impn}
\begin{split}
&E^{(1)}_{\mu\nu}=\frac{\partial\tilde{x}^\alpha}{\partial x^\mu}\frac{\partial\tilde{x}^\beta}{\partial x^\nu}\tilde{E}^{(1)}_{\alpha\beta}\\
\Rightarrow\,\,& E^{(1)}_{\tau i}=\exp\left(\frac{C^{(0)}}{2}\tau\right)\tilde{E}^{(1)}_{vi}-\rho\frac{C^{(0)}}{2}\exp\left(-\frac{C^{(0)}}{2}\tau\right)\tilde{E}^{(1)}_{ri}\\
\Rightarrow\,\,& E^{(1)}_{\tau i}|_{\rho=0}=\exp\left(\frac{C^{(0)}}{2}\tau\right)\tilde{E}^{(1)}_{vi}|_{r=0}\,.
\end{split}
\end{equation}
From the boost-weight arguments (see appendix \ref{app:boost}), we know that $\tilde{E}^{(1)}_{vi}$ carries a boost-weight of $+1$, {\it which in principle can arise either from a $\partial_v$ derivative or from the presence of $\tilde{A}_v$}. Now using \eqref{eq:imp1}, at the relevant order we are working here, $\tilde{A}_v$ has the following structure
\begin{equation}
\begin{split}
\tilde{A}_v^{(0)}&=r\,\chi_\tau^{(0)}\left(C^{(0)} rv/2,x^i\right)+\frac{C^{(0)}}{2}r A_\rho^{(0)}\left(C^{(0)} rv/2,x^i\right)\,,\\
&\equiv r\, Z^{(0)}(rv,x^i)\,.
\end{split}
\end{equation}
{\it Therefore, the boost-weight of $\tilde{A}_v$ is effectively carried by the overall factor of $r$. Although $r$ carries positive boost weight $+1$, it can not account for positive boost-weight when we are computing some quantities on the horizon $r=0$. Consequently, the positive boost weight $+1$ of $\tilde{E}^{(1)}_{vi}$ must be attributed to the derivative $\partial_v$.} Therefore all the boost-weight arguments will proceed as \cite{Bhattacharyya:2022nqa}.

Therefore, $\tilde{E}^{(1)}_{vi}$ contains one extra $\partial_v$ derivative with respect to $\partial_r$ or $\tilde{A}_r$. This extra $\partial_v$ acts on quantities that depend on $v$ through the product $rv$, therefore will bring a factor of $r$ which will vanish once evaluated on $r=0$ \footnote{if we want to construct a positive boost weight $+1$ term by acting $\partial_v$ on $\tilde{A}_r$, we have to act with two $\partial_v$ derivatives which will again bring an overall factor of $r$.}. Hence we conclude that $\tilde{E}^{(1)}_{vi}\Big|_{r=0}=0$. Therefore, from \eqref{eq:impn}, we obtain:

\begin{equation}\label{eq:imp3}
E^{(1)}_{\tau i}|_{\rho=0}=0\,.
\end{equation} 
Using this the total equation of motion at first subleading order becomes
\begin{equation}\label{eq:0th1storder}
\begin{split}
&E_{\tau i}^{(0)}\left[g^{(0)}_{\mu\nu}+\alpha\,g^{(1)}_{\mu\nu}, A^{(0)}_\mu+\alpha\,A_\mu^{(1)}\right]_{\rho=0}+\alpha\, E^{(1)}_{\tau i}\Big[g^{(0)}_{\mu\nu}\Big]_{\rho=0}\\
&=\alpha\,\partial_i X^{(1)}+\frac{1}{2}\alpha\,\partial_\alpha A_\tau^{(1)}F_{i}^{(0)\alpha}-\frac{1}{2}\alpha\, m^2 A^{(1)}_{\tau}A^{(0)}_{i}+\mathcal{O}(\alpha)^2\,.\\
&=\alpha\,\partial_i X^{(1)}+\frac{1}{2}\alpha\left(\partial_\rho A_\tau^{(0)}\right)\left(\partial_i A_\tau^{(1)}\right)+\frac{1}{2}\alpha\, h^{jk}\left(\partial_j A_\tau^{(1)}\right)F_{ik}^{(0)}-\frac{1}{2}\alpha\, m^2 A^{(1)}_{\tau}A^{(0)}_{i}+\mathcal{O}(\alpha)^2\,.\\
\end{split}
\end{equation}

We will now turn to the vector equation of motion $E_\mu$. Similar to the metric equation of motion, we can decompose $E_\mu$ into two parts : an {\it universal homogeneous part} and a {\it theory dependent source part}. Up to first subleading order it will be 
\begin{equation}\label{eq:vectorEoM}
E_\mu=E_\mu^{(0)}\left[g^{(0)}_{\mu\nu}+\alpha\,g^{(1)}_{\mu\nu}, A^{(0)}_\mu+\alpha\,A_\mu^{(1)}\right]+\alpha\,E_\mu^{(1)}\left[g^{(0)}_{\mu\nu}, A^{(0)}_\mu\right]\,.
\end{equation}
We have to analyse the $E_\tau$ component of the above equation at $\rho=0$. Let's first consider the homogeneous part
\begin{equation}\label{Etau1}
\begin{split}
&E_\tau^{(0)}\left[g^{(0)}_{\mu\nu}+\alpha\,g^{(1)}_{\mu\nu}, A^{(0)}_\mu+\alpha\,A_\mu^{(1)}\right]_{\rho=0}=\left(\alpha\,h^{ij}_{(0)}\mathcal{D}_i^{(0)}F^{(1)}_{j\tau}-\alpha\,m^2 A_\tau^{(1)}\right)_{\rho=0}
\end{split}
\end{equation}
where we have used the leading order equation of motion \eqref{eq:imp1} {\it i.e.,} $A_\tau^{(0)}=0$. See appendix \ref{app:Etau} (specifically eqn. \eqref{eq:Etau2nd}) for computational details.\\\\
Now, let's consider the source part
\begin{equation}\label{eq:sorc}
\alpha\,E_\tau^{(1)}\left[g^{(0)}_{\mu\nu}, A^{(0)}_\mu\right]_{\rho=0}\,.
\end{equation}
Using the boost-weight argument (see appendix \ref{app:boost}), we will demonstrate that the source term in equation \eqref{eq:sorc} vanishes. The reasoning closely parallels the derivation of equation \eqref{eq:imp3}, so we will not elaborate in details. Now, we will do the coordinate transformation \eqref{eq:coord_trans}. Under this coordinate transformation $E^{(1)}_\mu$ will transform as
\begin{equation}\label{eq:vectortransformq}
\begin{split}
E^{(1)}_\mu&=\frac{\partial \tilde{x}^\alpha}{\partial x^\mu} \tilde{E}^{(1)}_\alpha\\
\Rightarrow~E^{(1)}_\tau&=\exp\left(\frac{C^{(0)}}{2}\tau\right)\tilde{E}^{(1)}_v-\rho \frac{C^{(0)}}{2}\exp\left(-\frac{C^{(0)}}{2}\tau\right)\tilde{E}^{(1)}_r\\
\Rightarrow~E^{(1)}_\tau\Big|_{\rho=0}&=\exp\left(\frac{C^{(0)}}{2}\tau\right)\tilde{E}^{(1)}_v\Big|_{r=0}\,.
\end{split}
\end{equation}
Again, $\tilde{E}^{(1)}_v$ carries a positive boost weight of $+1$. Therefore, every term in $\tilde{E}^{(1)}_{v}$ includes an additional $\partial_v$ derivative. As a result, these terms vanish when evaluated at the horizon $\rho = 0$, as previously argued. Thus, by using the $\tau$-component of the vector equation of motion, $E_\tau = 0$, we obtain
\begin{equation}
\left(\alpha\,h^{ij}_{(0)}\mathcal{D}_i^{(0)}F^{(1)}_{j\tau}-\alpha\big[m^{(0)}\big]^2 A_\tau^{(1)}\right)_{\rho=0}=0\,.
\end{equation}
Now, we will use steps similar to \eqref{eq:step1} and \eqref{eq:step2}. Using stationarity condition we get
\begin{equation}\label{eq:stepp1}
F_{j\tau}^{(1)}=\partial_j A_\tau^{(1)}=\mathcal{D}_j^{(0)} A_\tau^{(1)}\,.
\end{equation}
We now multiply \eqref{eq:impn} with $\sqrt{h^{(0)}}A_\tau^{(1)}$ and integrate over a $D-2$ dimensional compact spatial manifold $C$ as we did in \eqref{eq:step2}
\begin{equation}\label{eq:stepp2}
\begin{split}
&\int_C d^{D-2}x \sqrt{h^{(0)}} A_\tau^{(1)}\left[h^{ij}_{(0)}\mathcal{D}^{(0)}_i \mathcal{D}^{(0)}_j A_\tau^{(1)}-\big[m^{(0)}\big]^2 A^{(1)}_\tau\right]_{\rho=0}=0\\
\Rightarrow&\int_C d^{D-2}x \sqrt{h^{(0)}} \left[-h^{ij}_{(0)}\left(\mathcal{D}^{(0)}_i A_\tau^{(1)}\right)\left(\mathcal{D}^{(0)}_j A_\tau^{(1)}\right)-m^2 \left(A^{(1)}_\tau\right)^2\right]_{\rho=0}=0\,.
\end{split}
\end{equation}
Since both terms are manifestly positive definite, they must vanish separately. This leads to the condition
\begin{equation}\label{eq:At1}
A_\tau^{(1)}\Big|_{\rho=0}=0\,.
\end{equation}
Using \eqref{eq:At1} in \eqref{eq:0th1storder} we finally get
\begin{equation}
\partial_i X^{(1)}\Big|_{\rho=0}=0\,,
\end{equation}
which completes the proof of the zeroth law at first subleading order. 

\subsection{Proof at arbitrary higher order}
In this subsection, we extend our proof to arbitrary orders in the effective field theory expansion using the {\it method of induction}. According to this method, if the proof holds at the leading order, and we can demonstrate that it holds at order $(n+1)$ assuming it holds at order $n$, then the proof is valid at all higher orders.

As we have already shown that the proof works at the leading order in subsection \ref{subsec:leading}, our focus here will be on establishing this inductive step $-$ namely, showing that the zeroth law holds at the $(n+1)$th order, provided it holds at the $n$th order. Strictly speaking, the proof at the first subleading order in subsection \ref{subsec:subleading} is not required for the induction to work. However, we include it as a clear and illustrative example of how the argument proceeds in practice. The essential framework required for generalization to higher perturbative orders is already embedded within the first subleading order proof. Therefore, we will keep this subsection brief.

Under our gauge choice, the metric takes the form given in equation \eqref{eq:metric}. For convenience, we explicitly write it out below:
\begin{equation}
\begin{split}
ds^2 = 2\, d\tau\, d\rho - \rho\, X(\rho, x^i)\, d\tau^2 + 2\rho\, \omega_i(\rho, x^i)\, d\tau\, dx^i + h{ij}(\rho, x^i)\, dx^i dx^j\,,
\end{split}
\end{equation}
where, $X$, $\omega_i$, $h_{ij}$ and also the vector field $A_\mu$ admit the effective field theory expansion written in equation \eqref{eq:gmunuexpansion}. We will assume that, we have solved both the equations of motion for the metric and also for the vector field up to order $\mathcal{O}(\alpha)^n$ implying
\begin{equation}\label{eq:assump1}
\begin{split}
&E_{\mu\nu}^{(0)}+\alpha\, E_{\mu\nu}^{(1)}+\cdots+\alpha^n E_{\mu\nu}^{(n)}=\mathcal{O}(\alpha)^{n+1}\,,\\
&E_{\mu}^{(0)}+\alpha\, E_{\mu}^{(1)}+\cdots+\alpha^n E_{\mu}^{(n)}=\mathcal{O}(\alpha)^{n+1}\,.\\
\end{split}
\end{equation}
We will assume that we have already shown that the extrinsic curvature is constant up to order $\mathcal{O}(\alpha)^n$ which implies
\begin{equation}\label{eq:condit}
\partial_i \left(X^{(0)}+\alpha\, X^{(1)}+\cdots+\alpha^n X^{(n)}\right)=\mathcal{O}(\alpha)^{n+1}\,.
\end{equation}

Now, we will consider the equations of motion at the order $\mathcal{O}(\alpha)^{n+1}$. As we have discussed at the beginning of section \ref{sec:proof}, we can divide the equations of motion into two parts $-$ homogeneous part and the source part. Let's first consider the homogeneous part. 
\begin{equation}
E^{(0)}_{\tau i}\left[g^{(0)}_{\mu\nu}+\alpha^{n+1}g_{\mu\nu}^{(n+1)},A^{(0)}_{\mu}+\alpha^{n+1}A_{\mu}^{(n+1)}\right]_{\rho=0}\,.
\end{equation}
Homogeneous part, by definition contains only and all the terms involving $g_{\mu\nu}^{(n+1)}$ and $A_\mu^{(n+1)}$. Therefore it has to have the same structure as that of \eqref{eq:E0taui} (see also appendix D of \cite{Bhattacharyya:2022nqa})

%\subsection{Case II: Temporal component of $A_\mu$ is dynamical}\label{subsec:caseII}
%In this subsection, we will consider a vector field with a dynamical temporal component, which may lead to ghost fields and other potential issues. However, since the pure gravity part also exhibits these problematic properties, we will proceed without paying attention to these issues.

\begin{equation}\label{eq:homoten}
\begin{split}
&E_{\tau i}^{(0)}\left[g^{(0)}_{\mu\nu}+\alpha^{n+1}\,g^{(n+1)}_{\mu\nu}, A^{(0)}_\mu+\alpha^{n+1}\,A_\mu^{(n+1)}\right]_{\rho=0}\\
&=\alpha^{n+1}\left[\partial_i X^{(n+1)}+\frac{1}{2}\left(\partial_\rho A_\tau^{(0)}\right)\left(\partial_i A_\tau^{(n+1)}\right)+\frac{1}{2} h^{jk}\left(\partial_j A_\tau^{(n+1)}\right)F_{ik}^{(0)}-\frac{1}{2} m^2 A^{(n+1)}_{\tau}A^{(0)}_{i}\right]+\mathcal{O}(\alpha)^{n+2}\,.\\
\end{split}
\end{equation}
Similarly, the homogeneous part of the vector equation has the following form \eqref{Etau1}
\begin{equation}\label{eq:homovec}
\begin{split}
&E_{\tau}^{(0)}\left[g^{(0)}_{\mu\nu}+\alpha^{n+1}\,g^{(n+1)}_{\mu\nu}, A^{(0)}_\mu+\alpha^{n+1}\,A_\mu^{(n+1)}\right]_{\rho=0}=\alpha^{n+1}\left(h^{ij}_{(0)}\mathcal{D}_i^{(0)}F^{(n+1)}_{j\tau}-m^2 A_\tau^{(n+1)}\right)_{\rho=0}+\mathcal{O}(\alpha)^{n+2}\,.\\
\end{split}
\end{equation}
We now turn to the computation of the source term. Before proceeding, we introduce a notation following \cite{Bhattacharyya:2022nqa}. Let's $Y^{(n)}$ denotes the coeffecient of $\alpha^n$ in the effective field theory expansion of $Y$. We will denote the series correct up to order $\mathcal{O}(\alpha)^n$ by 
\begin{equation}
Y^{[n]}=\sum_{k=0}^n\alpha^k Y^{(k)}\,.
\end{equation} 
In this notation, our assumption \eqref{eq:assump1} implies
\begin{equation}\label{eq:assump1p}
E_{\mu\nu}^{[n]}\left[g_{\mu\nu}^{[n]}, A_\mu^{[n]}\right]=\mathcal{O}(\alpha)^{n+1},\quad E_\mu^{[n]}\left[g_{\mu\nu}^{[n]}, A_\mu^{[n]}\right]=\mathcal{O}(\alpha)^{n+1}\,.
\end{equation}
The source parts in the $E_{\tau i}$ and $E_\tau$ equations of motion are respectively
\begin{equation}\label{eq:sourceppp}
E_{\tau i}^{[n+1]}\left[g_{\mu\nu}^{[n]},A_\mu^{[n]}\right]_{\rho=0},\quad \text{and,}\quad E_{\tau}^{[n+1]}\left[g_{\mu\nu}^{[n]},A_\mu^{[n]}\right]_{\rho=0},
\end{equation}
\eqref{eq:assump1p} implies both of \eqref{eq:sourceppp} are order $\mathcal{O}(\alpha)^{n+1}$ terms.\\\\
The truncated metric $g_{\mu\nu}^{[n]}$ has the following form
\begin{equation}
\begin{split}
ds^2 = 2\, d\tau\, d\rho - \rho\, X^{[n]}(\rho, x^i)\, d\tau^2 + 2\rho\, \omega^{[n]}_i(\rho, x^i)\, d\tau\, dx^i + h_{ij}^{[n]}(\rho, x^i)\, dx^i dx^j\,.
\end{split}
\end{equation}
The condition \eqref{eq:condit} implies
\begin{equation}
X^{[n]}(\rho,x^i)=C^{[n]}+\rho F^{[n]}(\rho,x^i)
\end{equation}
where $C^{[n]}$ is a constant and $F^{[n]}(\rho,x^i)$ denotes an arbitrary function of $\rho$ and $x^i$.

Now following subsection \ref{subsec:subleading}, we will do the following coordinate transformation from $x^\mu=\{\tau,\rho,x^i\}$ to $\tilde{x}^\mu=\{v,r,x^i\}$
\begin{equation}
\begin{split}
v=\frac{2}{C^{[n]}}\exp\left({\frac{C^{[n]}}{2}\tau}\right)\,, \quad r=\rho\,\exp\left(-{\frac{C^{[n]}}{2}\tau}\right)\,,
\end{split}
\end{equation}
where, $v$ is the affine parameter along the null generator of the Killing horizon.
The infinitesimal distance in this new coordinate take the following form
\begin{equation}\label{eq:aff_arb}
ds^2= 2dv\,dr-r^2 F^{[n]}\left(C^{[n]}rv/2\right)dv^2+r\,\omega_i^{[n]}\left(C^{[n]}rv/2\right)dv\, dx^i+h_{ij}^{[n]}\left(C^{[n]}rv/2\right) dx^i dx^j\,.
\end{equation}
Now, $E_{\mu\nu}^{[n]}$ and $E_\mu^{[n]}$ are covariant tensors of rank two and one, respectively. Hence, they transform as \eqref{eq:vectortransformq} and \eqref{eq:impn}
\begin{equation}
\begin{split}
E^{[n]}_\mu&=\frac{\partial \tilde{x}^\alpha}{\partial x^\mu} \tilde{E}^{[n]}_\alpha\\
\text{and,}\quad E^{[n]}_{\mu\nu}&=\frac{\partial\tilde{x}^\alpha}{\partial x^\mu}\frac{\partial\tilde{x}^\beta}{\partial x^\nu}\tilde{E}^{[n]}_{\alpha\beta}\,,
\end{split}
\end{equation}
which gives
\begin{equation}
\begin{split}
E^{[n]}_\tau\Big|_{\rho=0}&=\exp\left(\frac{C^{(0)}}{2}\tau\right)\tilde{E}^{[n]}_v\Big|_{r=0}\\
\text{and,}\quad E^{[n]}_{\tau i}|_{\rho=0}&=\exp\left(\frac{C^{(0)}}{2}\tau\right)\tilde{E}^{[n]}_{vi}|_{r=0}\,.
\end{split}
\end{equation}
Since both $\tilde{E}^{[n]}_v$ and $\tilde{E}^{[n]}_{vi}$ carry a positive boost weight of $+1$, the boost-weight argument from subsection \ref{subsec:subleading} (see also \ref{app:boost} for further details) implies that they must vanish. Consequently, we have
\begin{equation}
E^{[n]}_\tau\Big|_{\rho=0}=0,\quad \text{and,}\quad E^{[n]}_{\tau i}\Big|_{\rho=0}=0\,.
\end{equation}
The source terms in both equations of motion vanish, which implies that the homogeneous parts \eqref{eq:homovec} and \eqref{eq:homoten} must vanish independently. Vanishing of \eqref{eq:homovec} implies
\begin{equation}
\alpha^{n+1}\left(h^{ij}_{(0)}\mathcal{D}_i^{(0)}F^{(n+1)}_{j\tau}-m^2 A_\tau^{(n+1)}\right)_{\rho=0}=0\,.
\end{equation}
Using steps similar to \eqref{eq:stepp1} and \eqref{eq:stepp2} we get
\begin{equation}\label{eq:finall2}
A_\tau^{(n+1)}\Big|_{\rho=0}=0\,.
\end{equation}
Using \eqref{eq:finall2} in \eqref{eq:homoten} we get
\begin{equation}
\partial_i X^{(n+1)}\big|_{\rho=0}=0\,.
\end{equation}
This completes the proof that if we assume the constancy of surface gravity at order $\mathcal{O}(\alpha)^n$, then it also holds at order $\mathcal{O}(\alpha)^{n+1}$. Therefore, by the principle of {\it the method of induction}, the surface gravity remains constant to all orders in the effective field theory expansion.

\section{Some observations on more general vector theories}\label{sec:comments}
We now turn to a broader class of theories where gravity is coupled to a vector field $C_\mu$ that has all $D$ degrees of freedom propagating. Schematically, the effective field theory (EFT) action can be written like that in \eqref{eq:starting}.
%\begin{equation}
%S = \int d^D x \, \sqrt{-g} \;
%\Bigg( \sum_{n=2}^\infty \alpha^{\,n-2} \, \mathcal{L}_{(n)} \Bigg),
%\label{eq:generalEFT}
%\end{equation}
%a parameter constructed from the metric, curvature tensors, and the vector $C_\mu$.
  However, the leading-order Lagrangian $\mathcal{L}_{(0)}$ contains all possible kinetic terms that ensure $D$ propagating degrees of freedom \cite{deRham:2014zqa,Heisenberg:2014rta}. Explicitly, one may write
\begin{equation}\label{eq:Kinetic}
\mathcal{L}_{(0)} = R - \frac{1}{2} m^2 C_\mu C^\mu+c_1(\nabla_\mu C_\nu)(\nabla^\mu C^\nu)+c_2(\nabla_\mu C^\mu)(\nabla_\nu C^\nu)\,+c_3(\nabla_\mu C_\nu)(\nabla^\nu C^\mu)\,.
\end{equation}
This can be further modified to (for a detailed derivation, see Appendix \S\ref{app:non-proca})
\begin{equation}
\mathcal{L}_{(0)} = R + \frac{c_1}{2} \chi_{\mu\nu}\chi^{\mu\nu}
- \frac{1}{2} m^2 C_\mu C^\mu
+ (c_1+c_2+c_3) (\nabla_\mu C^\mu)^2
-(c_1+c_3) R_{\mu\nu} C^{\mu}C^\nu ,
\label{eq:L2general}
\end{equation}
with $\chi_{\mu\nu} = \nabla_\mu C_\nu - \nabla_\nu C_\mu$. From now on, let us set $c_1=-1/2$ so that, in analogy with the Maxwell case, the coefficient of the $\chi_{\mu\nu} \chi^{\mu\nu}$ term becomes $-1/4$.

Unlike the gauge-invariant case, these additional terms already appear at the same EFT order as the mass and Maxwell-like kinetic terms. Their presence generically renders \emph{all} $D$ components of $C_\mu$ dynamical, so that in four spacetime dimensions the theory typically propagates four degrees of freedom. In what follows, we examine the implications of such generalized vector theories for the zeroth law of black hole mechanics, and identify the restricted subclasses for which the law continues to hold consistently.

In order to examine the robustness of the zeroth law in this generalized setting, we adopt the following strategy: starting from the full Lagrangian \eqref{eq:L2general}, we selectively switch off certain terms and analyze 
the resulting field equations near the horizon. This procedure isolates the role of individual terms in enforcing the 
constraint $\partial_i X|_{\rho=0}=0$, which guarantees the constancy of the surface gravity. In particular, we focus on two cases.
% At the start of the section

\subsubsection*{Case I: $c_3=\frac{1}{2}$.}
 
In this case, theory reduces to the Einstein-Proca sector supplemented solely by the divergence term $(\nabla_\mu C^\mu)^2$
\begin{equation}
\mathcal{L}_{(0)} = R - \frac{1}{4} \chi_{\mu\nu}\chi^{\mu\nu} - \frac{1}{2} m^2 C_\mu C^\mu + c_2(\nabla_\mu C^\mu)^2 .
\label{eq:L2_caseI}
\end{equation}
Evaluating the $\tau i$ component of the corresponding field equations on the horizon (see Appendix~\ref{app:non-proca2}),  we get
\begin{equation}
E_{\tau i}^{(0)}|_{\rho=0} 
= R_{\tau i}^{(0)}
- \frac{1}{2} g^{\alpha\beta}\!\left(\chi_{\tau\alpha}^{(0)}\chi^{(0)}_{i \beta}{} 
\right)
-\frac{1}{2} m^2 C_\tau^{(0)}C_i^{(0)}
- c_2\,C_\tau^{(0)}\,\partial_i \!\big(\nabla\!\cdot\!C\big)^{(0)}\,.
\label{eq:mixEOM_caseI}
\end{equation}
Following the discussion in section \S \ref{sec:proof}, we now turn to the $\tau$ component of the vector equation of motion. Equation of motion for the vector field $C_{\mu}$ is
\begin{equation}
E^{(0)}_\mu =(\nabla^\nu \chi_{\nu\mu})^{(0)}-m^2 C^{(0)}_\mu\,
-2c_2 \nabla_\mu(\nabla_\nu C^\nu)^{(0)}\,,
\end{equation}
which implies
\begin{equation}
\begin{split}
E_{\tau}^{(0)} |_{\rho=0} 
&= (\,\nabla^\nu \chi_{\nu\tau} )^{(0)}
- m^2 C^{(0)}_\tau 
- 2c_2\,\nabla_\tau(\nabla\!\cdot\!C)^{(0)} ,\\&= \,h^{ij}_{(0)}\,\mathcal{D}^{(0)}_i \chi^{(0)}_{j\tau}
- m^2 C^{(0)}_\tau 
- 2c_2\cancel{\,\partial_\tau(\nabla\!\cdot\!C) }\,.
\label{eq:vec_tau_caseI}
\end{split}
\end{equation} 
The last term does not contribute, as the Killing vector for our gauge choice is $\partial_\tau$. Hence, we obtain an equation similar to the Einstein-Proca case. We multiply \eqref{eq:vec_tau_caseI} by$\sqrt{h}\,C_\tau^{(0)}$ and integrating over the compact horizon cross-section $C$, total 
derivatives drop out,  and we obtain a sum of squares
\begin{equation}
\int_C d^{D-2}x\,\sqrt{h^{(0)}}\,
\Big[\,-\, h_0^{ij}\, (\mathcal{D}_i C_\tau^{(0)})(\mathcal{D}_j C_\tau^{(0)}) - m^2 \big(C_\tau^{(0)}\big)^2 \Big]_{\rho=0}  = 0 ,
\label{eq:squares_caseI}
\end{equation}
implying,
\begin{equation}
C_\tau^{(0)}\Big|_{\rho=0}=0\,.
\label{eq:horizon_cond_caseI}
\end{equation}
Inserting \eqref{eq:horizon_cond_caseI} into \eqref{eq:mixEOM_caseI} leaves 
$E_{\tau i}^{(0)}|_{\rho=0} =\tfrac12\,\partial_i X^{(0)}$, so $E_{\tau i}^{(0)}=0$ implies
\[
\partial_i X^{(0)}\Big|_{\rho=0}=0 
\qquad\Longrightarrow\qquad 
\partial_i\kappa=0 .
\]
which establishes the constancy of the surface gravity on the horizon. Hence, the zeroth law continues to hold in this case.

\subsubsection*{Case II: $c_3=\text{arbitrary}$.}

Keeping the non–minimal coupling at leading order, the quadratic Lagrangian reads
\begin{equation}
\mathcal{L}_{(0)}
=  R
- \frac{1}{4} \chi_{\mu\nu}\chi^{\mu\nu}
- \frac{1}{2} m^2 C_\mu C^\mu
+ c'(\nabla_\mu C^\mu)^2
- d' R_{\mu\nu} C^\mu C^\nu .
\label{eq:L2_caseII}
\end{equation}
The mixed Einstein equation then takes the form (for detailed calculation see Appendix \ref{app:non-proca2}).
\begin{equation}\label{eq:L2_case}
\begin{split}
E_{\tau i}^{(0)}|_{\rho=0} &=-\frac{\partial_iX^{(0)}}{2}-\frac{1}{2}g^{\alpha\beta}\chi^{(0)}_{\tau\alpha}\chi^{(0)}_{i\beta}-\frac{1}{2} m^2 C^{(0)}_\tau C^{(0)}_{ i}\,-c^\prime C^{(0)}_\tau[\partial_i(\nabla.C)^{(0)}]-d^\prime  (2 C_i C^j R^{(0)}_{\tau j} + P C^{(0)}_\tau + Q \partial_i C^{(0)}_\tau ) \\
&=-\frac{\partial_iX^{(0)}}{2}-\frac{1}{2}g^{\alpha\beta}\chi^{(0)}_{\tau\alpha}\chi^{(0)}_{i\beta}-\frac{1}{2} m^2 C^{(0)}_\tau C^{(0)}_{ i}\,-c^\prime C^{(0)}_\tau [\partial_i(\nabla.C)^{(0)}]-d^\prime\left(-\partial_jX^{(0)} C_i C^j + P C^{(0)}_\tau + Q \partial_i C^{(0)}_\tau\right)\,.
\end{split}
 \end{equation}
Thus, for the zeroth law to hold, all terms except $\partial_i X^{(0)}$ in the above equation must vanish. This requires imposing the horizon conditions $C_\tau\Big|_{\rho=0}=0$. We try to derive this condition by manipulating the $\tau$-component of the vector equation of motion as we did previously. The equation of motion for the vector field is
\begin{equation}
E^{(0)}_\mu =(\nabla^\nu \chi_{\nu\mu})^{(0)}-m^2 C^{(0)}_\mu\,
-2c^{\prime} \nabla_\mu (\nabla_\nu C^\nu)^{(0)}
-2d^\prime R^{(0)}_{\mu\nu} C_{(0)}^\nu\,.
\end{equation}
Its $\tau $ component is
\begin{equation}
E^{(0)}_\tau |_{\rho=0} =
\,h^{ij}_{(0)}\,\mathcal{D}^{(0)}_i \chi^{(0)}_{j\tau}
- m^2 C^{(0)}_\tau
- 2 d^\prime \left[ 
C^{(0)}_\tau \,\mathcal{M}
- \tfrac{1}{2} h^{ij}\, \partial_i X^{(0)}\, C^{(0)}_j
\right],
\label{eq:vecTau_caseII}
\end{equation}
where
\begin{equation}
\mathcal{M}
= -\partial_\rho X^{(0)}- \tfrac{1}{2} \omega^{(0)}_i \omega_{(0)}^i
+ \tfrac{1}{2} \partial_i \omega^{i}_{(0)}
- \tfrac{1}{4}\partial_\rho h_{ij} h^{ij} X^{(0)}
+ \tfrac{1}{2} \Gamma^j_{jk}\,\omega^{k}_{(0)} .
\label{eq:calM_def}
\end{equation}
Multiplying \eqref{eq:vecTau_caseII} by $\sqrt{h}\,C_\tau^{(0)}$ and integrating over the compact cross-section $C$, we are unable to rearrange the integrand into whole square terms and therefore, we are not able to derive the required constraint on $C_\tau$. Thus, we could not establish the $\partial_i X^{(0)}$ in full generality for case~II. In short, fully dynamical vector fields $C_\mu$ differ qualitatively from gauge fields $B_\mu$ and massive Proca fields $A_\mu$, and demanding universal horizon thermodynamics for higher curvature theories of gravity coupled to vector fields $C_\mu$ might impose nontrivial constraints on admissible higher-derivative couplings.

\begin{comment}
the additional curvature term $R_{\mu\tau} C^\mu C_\tau$ obstructs the rearrangement of the integrand into a 
manifest sum of squares. In particular, the resulting integral cannot be reduced 
to a positive–definite expression (see Eqs.~(C.26)–(C.27)). As a consequence, the 
$\tau$–equation by itself does not enforce the conditions
\[
C_\tau\Big|_{\rho=0}=0, 
\qquad 
\partial_i C_\tau\Big|_{\rho=0}=0 .
\]

Substituting the undetermined \(C_\tau\) back into \eqref{eq:L2_case}, the $d'$–dependent source \(D_{\tau i}\) is generically non–vanishing on the horizon. Therefore, without imposing additional horizon constraints (such as \(C_\tau|_{\rho=0}=0\) and \(\partial_i C_\tau|_{\rho=0}=0\)). 

\end{comment}

\section{Conclusions and future directions}\label{sec:discussion}
In this note, we have shown that the surface gravity across a Killing horizon (not necessarily of bifurcate type) remains constant for stationary solutions of Einstein-Proca theory, including higher order corrections within a controlled EFT expansion. Since surface gravity can be identified with temperature, this result implies the constancy of temperature for such solutions in the presence of a Killing horizon $-$ precisely the statement of the zeroth law. In pure gravity, one demonstrates this by examining the $\tau i$ component of the equations of motion in a specific near horizon coordinate system \eqref{eq:metric}, as described in \cite{Bhattacharyya:2022nqa}. When a vector field is present, however, the $\tau i$ component alone is insufficient; one must also include the $\tau$ component of the vector equation of motion \cite{Davies:2024fut}. %This condition aligns with the generalized zeroth law, which shows that the constancy of surface gravity enforces $A_\tau|_{\rho=0}=\text{constant}$. (see Appendix~\ref{app:generalised} and \cite{Gao:2001ut,Prabhu:2015vua,Biswas:2022grc}).

The proof is manifestly perturbative: we expand the metric and the Proca field in the EFT parameter \eqref{eq:bhmetric} and isolate at each order of the EOM a universal homogeneous part and a theory-dependent source part. The homogeneous part is simple as it has the same universal structure at all orders, whereas the source part has different complicated forms in different perturbative orders as it depends on the details of the particular higher curvature terms. By applying “boost-weight” arguments, we have shown abstractly that the source term vanishes at each order (see \S\ref{sec:proof}). Implementing these arguments in the presence of a Proca field introduces subtleties beyond those encountered in the pure-gravity case \cite{Wall:2015raa,Bhattacharyya:2021jhr,Bhattacharyya:2022nqa}, which we now briefly discuss.

Consider the metric \eqref{eq:0thaff} and \eqref{eq:aff_arb} which are written in the horizon adapted coordinate system $\tilde{x}^\mu=\{v,r,x^i\}$ where $\partial_v$ is the null generator of the Killing horizon with $v$ being affine parameter. We write the metric here for ease of referencing in a slightly notationally convenient form
\begin{equation}
ds^2= 2dv\,dr-r^2 X\left(rv,x^i\right)dv^2+2\,r\,\omega_i\left(rv,x^i\right)dv\, dx^i+h_{ij}\left(rv,x^i\right) dx^i dx^j\,.
\end{equation}
This metric is invariant under the following boost-transformation as discussed in appendix \S\ref{app:boost}
\begin{equation}
r\rightarrow \lambda r,\quad v\rightarrow \frac{v}{\lambda}\,,
\end{equation}
where $\lambda$ is a constant. Then we can classify different terms according to their boost-weight, {\it e.g.}, if we consider $\tilde{E}_{vi}$ component of the equation of motion, it has boost-weight $+1$. In pure gravity, only $\partial_v$ carries positive boost-weight, which implies that on the Killing horizon ($r = 0$), $\tilde{E}_{vi}$ must include one extra $\partial_v$ compared to $\partial_r$. Therefore, for a stationary solution, since $\tilde{E}_{vi}$ depends on $r$ only through the product $r v$, it vanishes once evaluated at $r = 0$. This observation forms the core of the abstract argument presented in \cite{Bhattacharyya:2022nqa, Davies:2024fut} for the vanishing of the source term. In the presence of a Proca field, one might naively expect that the component $A_v$ itself can carry positive boost-weight, thereby invalidating the straightforward logic outlined above. However, we have shown that even with a Proca field, in the stationary scenario relevant for our discussion, it is only $\partial_v$ that carries positive boost weight (see Section \S\ref{sec:proof}).

In section \S\ref{sec:comments}, we have considered an even broader class of theories involving vector field $C_\mu$, where unlike the Proca theory, the temporal component of the vector field also propagates. For this general class, we are unable to prove the zeroth law for the most general Lagrangian, and succeeded only for a restricted subclass.
Our analysis of the two diagnostic cases discussed in section \S\ref{sec:comments} highlights the central role played by the curvature coupling in this generalized set up. In Case~I, where the non–minimal term 
$d' R_{\mu\nu} C^\mu C^\nu$ is absent, the zeroth law remains intact: the surface gravity 
is constant across the horizon once the standard regularity conditions are imposed. In Case II, however, where curvature coupling is retained at leading order, we are unable to demonstrate that $C_\tau=0$, and therefore cannot establish the zeroth law. %At present, a systematic characterization of the conditions under which the zeroth law can be restored in the presence of non–minimal curvature couplings is beyond reach within our framework.

Taken together, these results suggest that the zeroth law can serve as a practical diagnostic tool for constraining higher derivative terms. Requiring universal horizon thermodynamics may act as a filter, ruling out certain higher curvature contributions.  Pursuing this perspective more systematically $-$ potentially extending it to higher-spin EFTs $-$ is an important direction for future work. It would be interesting to investigate whether similar obstructions arise in massive spin-2 or higher-spin EFTs, and how these connect to positivity bounds and other consistency conditions.

It is also well known that these more general theories typically exhibit Ostrogradsky (ghost) instabilities, but certain special combinations of Lagrangian terms can cancel the ghost\cite{Heisenberg:2014rta}. It would therefore be interesting to investigate whether the zeroth law can be established under those ghost-free restrictions; we leave this question for future work.

As we have emphasized several times, our proof is inherently perturbative. A nonperturbative proof of the zeroth law would be a significant advancement and is left for future investigation.

%\newpage
%
%\begin{itemize}
%\item Working within the framework of EFT, we have showed that black holes in arbitrary higher derivative Proca theories satisfy the Zeroth Law of black hole thermodynamics. 
%\item To achieve this, it was necessary to analyze the off-shell structure of $E_{vi}$ $-$ similar to \cite{Wall:2024lbd}.
%\item The $\tau i$ component of the equations of motion alone is insufficient; the $\tau$ component of the vector equation of motion must also be considered.
%\item Write some future directions regarding non-Proca vector theories.
%\item Write some future direction in the line of \cite{Hollands:2022ajj}
%\end{itemize}

\section*{Acknowledgements}
We are grateful to Sayantani Bhattacharyya and Nilay Kundu for numerous insightful discussions throughout the course of this work, and for going through the draft, offering critical feedback that significantly enhanced the quality of the paper. We thank Iain Davies, Bobby Ezhuthachan, Suman Kundu, Mangesh Mandlik and Binata Panda for valuable discussions. PB gratefully acknowledges the support provided by the grant CRG/2021/004539 during the initial stages of this work. We also acknowledge the ST4 2025 program at IISER Bhopal and the National Strings Meeting 2024 at IIT Ropar, 
both of which provided stimulating environments where parts of this work were discussed and refined. 
Finally, we express our deep appreciation to the people of India for their continued support of basic scientific research.

\appendix
\section{Some computational details}\label{app:AA}
\subsection{Inverse metric, Christoffel symbols and Ricci tensor}\label{app:details1}
In this appendix we will compute Christoffel symbols and Ricci tensor for the metric
\begin{equation}
ds^2=2\, d\tau\, d\rho-\rho\, X(\rho,x^i)d\tau^2+2\,\rho\,\omega_i(\rho,x^i)d\tau dx^i+h_{ij}(\rho,x^i)dx^i dx^j
\end{equation}
%Different components of the metric are
%\begin{equation}
%\begin{split}
%&g_{\tau\tau}=-\rho X(\rho,x^i),~~g_{\tau\rho}=1,~~g_{\tau i}=\rho~ \omega_i(\rho,x^i),\\
%&g_{\rho\rho}=0,~~g_{\rho i}=0,~~g_{ij}=h_{ij}(\rho,x^i)
%\end{split}
%\end{equation}
The inverse metric components are
\begin{equation}
\begin{split}
&g^{\tau\tau}=0,~~g^{ \tau\rho}=1,~~g^{\tau i}=0,~~g^{\rho\rho}=\rho~X+\rho^{2}\omega^2 \\
&g^{\rho i}=-\rho~\omega^{i},~~ g^{ij}=h^{ij}
\end{split}
\end{equation}
where, $h^{ij}$ is defined as $h^{ik}h_{kj}=\delta^i_j$, and  $\omega^{i}$ is defined as $\omega^{i}=h^{ij}\omega_j$, and $\omega^2$ is defined as $\omega^2=\omega^i \omega_i$\\
\\
We now proceed to the computation of Christoffel symbols. Different components of the Christoffel symbols on the horizon ($\rho=0$) are 
\begin{equation}
\begin{split}
\Gamma^{\rho}_{\rho\rho}&=0, \quad \Gamma^{\rho}_{\tau\rho}=-\frac{1}{2}X, \quad \Gamma^{\rho}_{\rho j}=\frac{1}{2}\omega_{j},\quad \Gamma^{\rho}_{i\tau}=0, \quad \Gamma^\rho_{ij}=0, \quad \Gamma^\rho_{\tau\tau}=0,\quad\Gamma^\tau_{\tau\tau}=\frac{1}{2}X,\quad \Gamma^\tau_{\tau\rho}=0,\\
\Gamma^\tau_{\rho\rho}&=0,\quad \Gamma^\tau_{\tau i}=-\frac{1}{2}\omega_i,\quad \Gamma^\tau_{ij}=-\frac{1}{2}\partial_\rho h_{ij},\quad\Gamma^\tau_{i\rho}=0, \quad \Gamma^j_{i\tau}=0,\quad \Gamma^j_{\tau\tau}=0,\quad \Gamma^i_{j\rho}=\frac{1}{2}h^{ik}\partial_\rho h_{jk},\\
\Gamma^i_{jk}&\equiv \hat{\Gamma}^i_{jk}=\frac{1}{2}h^{im}\left(\partial_j h_{km}+\partial_k h_{jm}-\partial_m h_{jk}\right),\quad\Gamma^i_{\tau\rho}=\frac{1}{2}h^{ij}\omega_j,\quad \Gamma^i_{\rho\rho}=0\,.
\end{split}
\end{equation}
We will require the following components of the Christoffel symbols off the horizon upto order $\mathcal{O}(\rho)$
\begin{equation}
\begin{split}
\Gamma^{\rho}_{i\tau}&=-\frac{\rho}{2}\left[X\omega_{i}+\partial_{i}X\right],\quad \Gamma^\tau_{\tau i}=-\frac{1}{2}\left[\omega_i+\rho\partial_\rho\omega_i\right],\quad \Gamma^j_{i\tau}=\frac{1}{2}\rho\omega^j\omega_i-\frac{\rho}{2}h^{jk}\left[\partial_k \omega_i-\partial_i\omega_k\right]\,,\\
\Gamma^\tau_{\tau\tau}&=\frac{1}{2}\left[X+\rho\partial_\rho X\right],\quad \Gamma^\rho_{\tau\tau}=\frac{1}{2}\rho X^2,\quad \Gamma^j_{\tau\tau}=-\frac{\rho}{2}X h^{jk}\omega_k+\frac{\rho}{2}h^{jk}\partial_k X
\end{split}
\end{equation}
With this, we can compute Ricci tensors, relevant for our analysis. \\\\
Evaluation of $R_{\tau i}$ on the horizon
\begin{equation}
R_{\tau i}={R^\tau}_{\tau\tau i}+{R^\rho}_{\tau\rho i}+{R^j}_{\tau j i}
\end{equation}
%Let's first compute different components of the Riemann tensor
\begin{equation}
\begin{split}
{R^\tau}_{\tau\tau i}\Big|_{\rho=0}&=\partial_\tau \Gamma^\tau_{i\tau}-\partial_i\Gamma^\tau_{\tau\tau}+\Gamma^\tau_{\tau \lambda}\Gamma^\lambda_{i\tau}-\Gamma^\tau_{i\lambda}\Gamma^\lambda_{\tau\tau}\,,\\
&=-\frac{1}{2}\partial_i X\,.
\end{split}
\end{equation}
\begin{equation}
\begin{split}
{R^\rho}_{\tau\rho i}\Big|_{\rho=0}&=\partial_\rho \Gamma^\rho_{i\tau}-\partial_i\Gamma^\rho_{\rho\tau}+\Gamma^\rho_{\rho \lambda}\Gamma^\lambda_{i\tau}-\Gamma^\rho_{i\lambda}\Gamma^\lambda_{\rho\tau}\\
&=0
\end{split}
\end{equation}
\begin{equation}
\begin{split}
{R^j}_{\tau j i}\Big|_{\rho=0}&=\partial_j \Gamma^j_{i\tau}-\partial_i\Gamma^j_{j\tau}+\Gamma^j_{j\lambda}\Gamma^\lambda_{i\tau}-\Gamma^j_{i\lambda}\Gamma^\lambda_{j\tau}\\
&=0
\end{split}
\end{equation}
Finally we get
\begin{equation}\label{eq:Rtaui}
R_{\tau i}\Big|_{\rho=0}=-\frac{1}{2}\partial_i X\Big|_{\rho=0}
\end{equation}
Evaluation of $R_{\tau\rho}$ on the horizon,
\begin{equation}
R_{\tau \rho}={R^\tau}_{\tau\tau \rho}+{R^\rho}_{\tau\rho \rho}+{R^j}_{\tau j \rho}
\end{equation}
Let's first compute different components of the Riemann tensor
\begin{equation}
\begin{split}
{R^\tau}_{\tau\tau \rho}\Big|_{\rho=0}&=\partial_\tau \Gamma^\tau_{\rho\tau}-\partial_\rho\Gamma^\tau_{\tau\tau}+\Gamma^\tau_{\tau \lambda}\Gamma^\lambda_{\rho\tau}-\Gamma^\tau_{\rho\lambda}\Gamma^\lambda_{\tau\tau}\,,\\
&=-\partial_\rho X\,-\frac{1}{4} \omega_i \omega^i.
\end{split}
\end{equation}
\begin{equation}
\begin{split}
{R^\rho}_{\tau\rho \rho}\Big|_{\rho=0}&=\partial_\rho \Gamma^\rho_{\rho\tau}-\partial_\rho\Gamma^\rho_{\rho\tau}+\Gamma^\rho_{\rho \lambda}\Gamma^\lambda_{\rho\tau}-\Gamma^\rho_{\rho\lambda}\Gamma^\lambda_{\rho\tau}\\
&=0
\end{split}
\end{equation}
\begin{equation}
\begin{split}
{R^j}_{\tau j \rho}\Big|_{\rho=0}&=\partial_j \Gamma^j_{\rho\tau}-\partial_\rho\Gamma^j_{j\tau}+\Gamma^j_{j\lambda}\Gamma^\lambda_{\rho\tau}-\Gamma^j_{\rho\lambda}\Gamma^\lambda_{j\tau}\\
&=-\frac{1}{4} \omega_i \omega^i-\frac{1}{4} \partial_\rho h_{ij} h^{ij} X +\frac{1}{2} \partial_i \omega^i +\frac{1}{2} \Gamma^j_{jk} \omega^k
\end{split}
\end{equation}
Finally we get
\begin{equation}\label{eq:Rtaurho}
R_{\tau \rho}\Big|_{\rho=0}=-\partial_\rho X-\frac{1}{2} \omega^i \omega_i -\frac{1}{4} \partial_\rho h_{ij} h^{ij} X +\frac{1}{2} \partial_i \omega^i +\frac{1}{2} \Gamma^j_{jk} \omega^k
\end{equation}
In this appendix, we have thus far refrained from performing any expansion in the effective field theory parameter $\alpha$. In principle, any of the expressions above can be expanded to the desired order. For instance, let us compute $h^{ij}$ up to order $\mathcal{O}(\alpha)^2$. To do this we have to expand $X(\rho,x^i), \omega_i(\rho,x^i)$ and $h_{ij}(\rho,x^i)$ up to order $\mathcal{O}(\alpha)^2$.
\begin{equation}
\begin{split}
X(\rho,x^i)&=X^{(0)}(\rho,x^i)+\alpha\, X^{(1)}(\rho,x^i)+\alpha^2 X^{(2)}(\rho, x^i)\,,\\
\omega_i(\rho,x^i)&=\omega_i^{(0)}(\rho,x^i)+\alpha\,\omega_i^{(1)}(\rho,x^i)+\alpha^2 \omega_i^{(2)}(\rho,x^i)\,,\\
h_{ij}(\rho,x^i)&=h_{ij}^{(0)}(\rho,x^i)+\alpha\,h_{ij}^{(1)}(\rho,x^i)+\alpha^2 h_{ij}^{(2)}(\rho,x^i)\,.
\end{split}
\end{equation}
Then we can compute $h^{ij}$ which has the following expression
\begin{equation}
h^{ij}=\big[h^{(0)}\big]^{ij}+\alpha \big[h^{(1)}\big]^{ij}+\alpha^2 \big[h^{(2)}\big]^{ij}
\end{equation}
where, $\big[h^{(0)}\big]^{ij}$ is the inverse of $h_{ij}^{(0)}$ and
\begin{equation}
\begin{split}
\big[h^{(1)}\big]^{ij}&=-\big[h^{(0)}\big]^{im}h^{(1)}_{mn}\big[h^{(0)}\big]^{nj}\,,\\
\big[h^{(2)}\big]^{ij}&=\big[h^{(0)}\big]^{im}h^{(1)}_{mk}\big[h^{(0)}\big]^{k\ell}h^{(1)}_{\ell n}\big[h^{(0)}\big]^{nj}-\big[h^{(0)}\big]^{im}h^{(2)}_{mn}\big[h^{(0)}\big]^{nj}\,.
\end{split}
\end{equation}
In a similar manner, we can compute other quantities as well. Next, we provide the expression of $\hat{\Gamma}^i_{jk}$ up to order $\mathcal{O}(\alpha)$, as this will be necessary for our further computation
\begin{equation}
\hat{\Gamma}^i_{jk}=[\hat{\Gamma}^{(0)}]^i_{jk}+\alpha\, [{\hat{\Gamma}^{(1)}}]^i_{jk}\,,
\end{equation}
where,
\begin{equation}
\begin{split}
[\hat{\Gamma}^{(0)}]^i_{jk}&=\frac{1}{2}h^{i\ell}_{(0)}\left[\partial_j h_{k\ell}^{(0)}+\partial_k h^{(0)}_{j\ell}-\partial_\ell h_{jk}^{(0)}\right]\\
[\hat{\Gamma}^{(1)}]^i_{jk}&=\frac{1}{2}\,h^{i\ell}_{(0)}\left[\partial_j h_{k\ell}^{(1)}+\partial_k h^{(1)}_{j\ell}-\partial_\ell h_{jk}^{(1)}\right]-\frac{1}{2} h^{i\ell}_{(1)}\left[\partial_j h_{k\ell}^{(0)}+\partial_k h^{(0)}_{j\ell}-\partial_\ell h_{jk}^{(0)}\right]\,.
\end{split}
\end{equation}
%%
%%%
From \eqref{eq:Rtaui} we can write $R_{\tau i}\Big|_{\rho=0}$ up to any arbitrary order as
\begin{equation}
R_{\tau i}\Big|_{\rho=0}=-\frac{1}{2}\partial_i\Big(X^{(0)}(\rho,x^j)+\alpha\, X^{(1)}(\rho,x^j)+\alpha^2 X^{(2)}(\rho,x^j)+\cdots\Big)_{\rho=0}
\end{equation}

\subsection{Simplification of $E_\tau^{(0)}$}\label{app:Etau}
$E^{(0)}_\mu$ has the following structure
\begin{equation}
E^{(0)}_{\mu}=\nabla^\nu F_{\nu\mu}-m^2 A_\mu
\end{equation}
In this appendix, we will explicitly compute $E^{(0)}_{\tau}\left[g^{(0)}_{\alpha\beta}+\alpha\, g^{(1)}_{\alpha\beta}, A^{(0)}_\alpha+\alpha\, A_\alpha^{(1)}\right]$ upto $\mathcal{O}(\alpha)$. 
\begin{equation}
\begin{split}
E_\tau^{(0)}&=\nabla^\nu F_{\nu\tau}-m^2 A_\tau\\
&=g^{\nu\lambda}\nabla_\lambda F_{\nu\tau}-m^2 A_\tau
\end{split}
\end{equation}
Here, the coordinate system is $x^\mu=\{\tau, \rho, x^i\}$. Therefore, we get
\begin{equation}
E^{(0)}_\tau=g^{\rho\lambda}\nabla_\lambda F_{\rho\tau}+g^{i\lambda}\nabla_\lambda F_{i\tau}-m^2 A_\tau
\end{equation}
Now, we evaluate the above equation at $\rho=0$.
\begin{equation}
\begin{split}
E^{(0)}_\tau\Big|_{\rho=0}&=\left(\nabla_\tau F_{\rho\tau}+h^{ij}\nabla_j F_{i\tau}-m^2 A_\tau\right)\Big|_{\rho=0}\\
&=\Big[\cancel{\partial_\tau F_{\rho\tau}}-\Gamma^\lambda_{\tau\rho}F_{\lambda\tau}-\Gamma^\lambda_{\tau\tau}F_{\rho\lambda}+h^{ij}\left(\partial_j F_{i\tau}-\Gamma^\lambda_{ji}F_{\lambda\tau}-\Gamma^\lambda_{j\tau}F_{i\lambda}\right)-m^2 A_\tau\Big]_{\rho=0}\\
&=\Big[-\Gamma^\rho_{\tau\rho}F_{\rho\tau}-\Gamma^i_{\tau\rho}F_{i\tau}-\Gamma^\tau_{\tau\tau}F_{\rho\tau}+h^{ij}\left(\partial_j F_{i\tau}-\hat{\Gamma}^k_{ji}F_{k\tau}-\Gamma^\tau_{j\tau}F_{i\tau}\right)-m^2 A_\tau\Big]_{\rho=0}\\
&=\Big[-\Gamma^i_{\tau\rho}F_{i\tau}+h^{ij}\left(\partial_j F_{i\tau}-\hat{\Gamma}^k_{ji}F_{k\tau}-\Gamma^\tau_{j\tau}F_{i\tau}\right)-m^2 A_\tau\Big]_{\rho=0}\\
&=\Big[h^{ij}\left(\partial_j F_{i\tau}-\hat{\Gamma}^k_{ji}F_{k\tau}\right)-m^2 A_\tau\Big]_{\rho=0}
\end{split}
\end{equation}
Now, we will use EFT expansion for the metric and the vector field and expand $E_\tau^{(0)}$ in powers of $\alpha$. Schematically, we represent this as
\begin{equation}
E_{\tau}^{(0)}\left[g^{(0)}_{\alpha\beta}+\alpha\,g^{(1)}_{\alpha\beta}+\cdots, A_\alpha^{(0)}+\alpha\, A_\alpha^{(1)}+\cdots\right]_{\rho=0}=\left(E_{\tau}^{(0,0)}+\alpha\, E_{\tau}^{(0,1)}+\alpha^2 E_{\tau}^{(0,2)}+\cdots\right)_{\rho=0}
\end{equation}
Now, we will explicitly compute $E_\tau^{(0)}$ up to order $\mathcal{O}(\alpha)$
\begin{equation}
\begin{split}
&E_{\tau}^{(0)}\left[g^{(0)}_{\alpha\beta}+\alpha\,g^{(1)}_{\alpha\beta}+\cdots, A_\alpha^{(0)}+\alpha\, A_\alpha^{(1)}+\cdots\right]_{\rho=0}\\
&=\left[\big[h^{(0)}\big]^{ij}+\alpha \big[h^{(1)}\big]^{ij}\right]\left[\partial_j F_{i\tau}^{(0)}+\alpha\, \partial_j F_{i\tau}^{(1)}-\left([\hat{\Gamma}^{(0)}]^i_{jk}+\alpha\, [{\hat{\Gamma}^{(1)}}]^i_{jk}\right)\left(F_{k\tau}^{(0)}+\alpha F_{k\tau}^{(1)}\right)\right]_{\rho=0}\\
&~~~~-\left(m^{(0)}+\alpha\, m^{(1)}\right)^2\left(A_\tau^{(0)}+\alpha A_\tau^{(1)}\right)\\
&=\big[h^{(0)}\big]^{ij}\left[\partial_j F_{i\tau}^{(0)}-[\hat{\Gamma}^{(0)}]^i_{jk}F_{k\tau}^{(0)}\right]_{\rho=0}-\big[m^{(0)}\big]^2 A_\tau^{(0)}\\
&+\alpha\,\big[h^{(1)}\big]^{ij}\left[\partial_j F_{i\tau}^{(0)}-[\hat{\Gamma}^{(0)}]^i_{jk}F_{k\tau}^{(0)}\right]_{\rho=0}+\alpha\big[h^{(0)}\big]^{ij}\left(\partial_j F_{i\tau}^{(1)}-[{\hat{\Gamma}^{(0)}}]^k_{ji}F^{(1)}_{k\tau}-[{\hat{\Gamma}^{(1)}}]^k_{ji}F^{(0)}_{k\tau}\right)\\
&-2\,\alpha\, m^{(0)}m^{(1)}A_\tau^{(0)}-\alpha \big[m^{(0)}\big]^2 A_\tau^{(1)}
\end{split}
\end{equation}
Therefore we finally get
\begin{equation}\label{eq:Etau2nd}
\begin{split}
E_\tau^{(0,0)}&=\big[h^{(0)}\big]^{ij}\left[\partial_j F_{i\tau}^{(0)}-[\hat{\Gamma}^{(0)}]^i_{jk}F_{k\tau}^{(0)}\right]_{\rho=0}-\big[m^{(0)}\big]^2 A_\tau^{(0)}\,,\\
&=\big[h^{(0)}\big]^{ij}\mathcal{D}^{(0)}_j F^{(0)}_{i\tau}-\big[m^{(0)}\big]^2 A_\tau^{(0)}\,.\\
E_\tau^{(0,1)}&=\big[h^{(1)}\big]^{ij}\left[\partial_j F_{i\tau}^{(0)}-[\hat{\Gamma}^{(0)}]^i_{jk}F_{k\tau}^{(0)}\right]_{\rho=0}+\big[h^{(0)}\big]^{ij}\left(\partial_j F_{i\tau}^{(1)}-[{\hat{\Gamma}^{(0)}}]^k_{ji}F^{(1)}_{k\tau}-[{\hat{\Gamma}^{(1)}}]^k_{ji}F^{(0)}_{k\tau}\right)\\
&~~~~-2\, m^{(0)}m^{(1)}A_\tau^{(0)}-\big[m^{(0)}\big]^2 A_\tau^{(1)}\,,\\
&=\big[h^{(1)}\big]^{ij}\mathcal{D}^{(0)}_j F^{(0)}_{i\tau}+\big[h^{(0)}\big]^{ij}\mathcal{D}^{(0)}_j F^{(1)}_{i\tau}-\big[h^{(0)}\big]^{ij}[{\hat{\Gamma}^{(1)}}]^k_{ji}F^{(0)}_{k\tau}-2\, m^{(0)}m^{(1)}A_\tau^{(0)}-\big[m^{(0)}\big]^2 A_\tau^{(1)}\,,\\
\end{split}
\end{equation}
where, $\mathcal{D}^{(0)}_j$ is covariant derivative with respect to $h_{ij}^{(0)}$.

\section{Summary of boost-weight arguments}\label{app:boost}

The most general near-horizon metric of a stationary black hole is given by \ref{eq:metric}. If the zeroth law is satisfied, that is $\partial_i X|_{\rho=0}=0$, we can obtain a general solution of the metric coefficient as $ X|_{\rho=0}$.
\begin{equation}
    X(\rho,x^i)=C+ \rho~F(\rho,x^i),
\end{equation}
where $C$ is an integration constant and $F(\rho,x^i)$ is an arbitrary function of $(\rho,x^i)$. The condition $\partial_i X|_{\rho=0}$ fixes $C=2 \kappa$. With this \eqref{eq:metric} becomes
\begin{equation}
\begin{split}
ds^2&=2 d\tau~ d\rho-\rho \left(C+\rho\,F(\rho,x^i)\right) d\tau^2+2 \rho~ \omega_i(\rho,x^i) d\tau d x^i+h_{ij}(\rho,x^i) dx^i dx^j\,.
\end{split}
\end{equation}
Here, we can perform a particular coordinate transformation from $(\rho,\tau,x^i)$ to $(r,v,x^i)$ by
\begin{equation}
\begin{split}
v=\frac{2}{C}\exp\left({\frac{C}{2}\tau}\right)\,,\\
r=\rho\,\exp\left(-{\frac{C}{2}\tau}\right)\,,
\end{split}
\end{equation} to obtain
\begin{equation}\label{eq:rv1}
ds^2= 2dv\,dr-r^2 F\left(C rv/2\right)dv^2+r\,\omega_i\left(C rv/2\right)dv\, dx^i+h_{ij}\left(C rv/2\right) dx^i dx^j\,.
\end{equation}
The horizon is now located at $r=0$. This choice of metric is different from the previous one because here $\partial_v$ is the affinely parameterised null generator of the horizon. In the metric \eqref{eq:metric}, the metric components were independent of the coordinate $\tau$, but here in \eqref{eq:rv1}, they are functions of $v$, specifically functions of the product $rv$. This form of the metric is useful because, despite fixing the gauge, there is still some residual freedom to perform a scaling transformation of the following form: (For a more detailed discussion, refer to \S 3 of\cite{Biswas:2022grc})

\begin{equation}
    r \rightarrow \lambda r~~~~ \text{along with } v \rightarrow\frac{v}{\lambda},
\end{equation}
where $\lambda$ is a constant parameter. This particular transformation keeps \eqref{eq:rv1}, invariant, which we call boost symmetry. This information about the stationary black hole metric 
helps us to understand how any covariant tensor will transform under this boost transformation. Therefore, the metric components in a stationary or equilibrium configuration will be of the form 
$g^{bh}_{\mu\nu}(rv,x^i)$, because it respects the boost symmetry. Any quantity outside this function is out of equilibrium, and the boost weight explains their stationarity deviation. 
Any covariant tensor $\mathcal{A}$ will transform in the following way
\begin{equation}\label{eq:bosstwt}
\mathcal{A}\rightarrow \tilde{\mathcal{A}}=\lambda^w \mathcal{A},  
\end{equation}
where $w$ is the boost weight of $\mathcal{A}$. $\mathcal{A}$ is positive boost weight quantity if 
\begin{equation}
    A(rv,x^i) \rightarrow \partial^{m_r}_r \partial^{m_v}_v \tilde{A}(rv,x^i)|_{r=0}~~ \text{whenever}~~~~ m_v> m_r\,.
\end{equation}
$\mathcal{A}$ being a positive boost weight quantity of boost weight $m_v-m_r$ means if one operates $\partial^{m_r}_r \partial^{m_v}_v$  on $\tilde{A}(rv,x^i)$ ,it will leave behind an extra $r$, and hence it will vanish on the horizon $r=0$. Thus, positive boost weight quantities will vanish on a stationary horizon. Thus, we can use this useful result if we can write in this coordinate system \eqref{eq:rv1}. Similar discussions have been done for the case of gauged vector fields in \S 4 of \cite{Biswas:2022grc}.

\section{Calculations of more general vector theories}\label{app:nonProcaL}
\subsection{Derivation of the general non-Proca Lagrangian}\label{app:non-proca}

The leading-order terms in the most general higher-curvature gravity 
theories with non-minimal couplings to vector fields, 
particularly non-gauge vector fields, can be represented by
\begin{equation}\label{eq:nplagrangian}
\begin{split}
 \mathcal{L}_{(0)}&=R- \tfrac{1}{2} m^2 C_\mu C^\mu 
 + c_1 (\nabla_\mu C_\nu)(\nabla^\mu C^\nu)
 + c_2 (\nabla_\mu C^\mu)(\nabla_\nu C^\nu)
 + c_3 (\nabla_\mu C_\nu)(\nabla^\nu C^\mu) .
\end{split}
\end{equation}
The various derivative structures can be simplified. 
For instance,
\begin{equation}
c_1 (\nabla_\mu C_\nu)(\nabla^\mu C^\nu) 
= \tfrac{c_1}{2} \chi_{\mu\nu}\chi^{\mu\nu}
+ c_1 \nabla^\mu C_\nu \nabla^\nu C_\mu ,
\end{equation}
while
\begin{equation}
c_3 (\nabla_\mu C_\nu)(\nabla^\nu C^\mu) 
= c_3 (\nabla_\mu C^\mu)(\nabla_\nu C^\nu) 
- c_3 C^\mu R_{\mu\nu} C^\nu.
\end{equation}
Combining these relations and absorbing coefficients into redefinitions, 
the final form of the Lagrangian can be written as
\begin{equation}
\begin{split}
\mathcal{L}_{(0)}&=R
+ \tfrac{c_1}{2} \chi_{\mu\nu}\chi^{\mu\nu}
-\frac{1}{2}m^2 C_\mu C^\mu
+ (c_1+c_2+c_3) (\nabla_\mu C^\mu)^2
- (c_1+c_3) R_{\mu\nu} C^\mu C^\nu \\
\end{split}
\label{eq:appendixfinal}
\end{equation}
which is the compact form quoted in Eq.~\eqref{eq:L2general} of the main text. 

%In particular, choosing $c_1=-\tfrac{1}{2}$ and $b_1=-1$ 
%reduces \ref{eq:appendixfinal} to the Lagrangian used in the main discussion. This choice of coefficients will still keep $D$ degrees of freedom propagating.

\subsection{Computational details for case I}\label{app:non-proca2}
\begin{equation}
\mathcal{L}_{(0)} = R-\frac{1}{4}\chi_{\mu\nu}\chi^{\mu\nu}-\frac{1}{2}m^2 C_\mu C^\mu+ c_2 (\nabla_\mu C^\mu)^2 .
\end{equation}
Equation of motion w.r.t. the metric $g_{\mu\nu}$
\begin{equation}\label{eq:eoms2}
\begin{split}
E_{\mu\nu}^{(0)}&=R_{\mu\nu}-\frac{1}{2}g_{\mu\nu}R+\frac{1}{8}g_{\mu\nu} \chi_{\alpha \beta} \chi^{\alpha \beta}-\frac{1}{2}g^{\alpha\beta}\chi_{\mu\alpha}\chi_{\nu\beta}+\frac{1}{4}m^2g_{\mu\nu} C^\alpha C_\alpha-\frac{1}{2} m^2 C_{\mu}C_{\nu}+c_2 H_{\mu\nu}\,,
\end{split}
\end{equation}
where,
\begin{equation}\label{eq:cmunu}
\begin{split}
H^{(0)}_{\mu\nu}&=-\frac{1}{2} g_{\mu \nu}\,\nabla_\alpha C^\alpha \nabla_\beta C^\beta +(\nabla_\mu C_\nu+\nabla_\nu C_\mu) \nabla_\alpha A^\alpha-\nabla_\mu\left(C_\nu\nabla_\alpha C^\alpha\right)-\nabla_\nu\left(C_\mu\nabla_\beta C^\beta\right)\\&+g_{\nu\mu}\nabla^\sigma\left(C_\sigma\nabla_\beta C^\beta \right)\,.
\end{split}
\end{equation}
Equation of motion w.r.t. vector field $C_{\mu}$
\begin{equation}
E^{(0)}_\mu =\nabla^\nu \chi_{\nu\mu}-m^2 C_\mu\,
-2c_2 \nabla_\mu(\nabla_\nu C^\nu).
\end{equation}
For the zeroth law, the relevant component of the EOM is
the $E_{\tau i}$ component,
\begin{equation}\label{eq:etaui2}
\begin{split}
E_{\tau i}^{(0)}&=R_{\tau i}-\frac{1}{2}g_{\tau i}R+\frac{1}{8}g_{\tau i} \chi_{\alpha \beta} \chi^{\alpha \beta}-\frac{1}{2}g^{\alpha\beta}\chi_{\tau\alpha}\chi_{i\beta}+\frac{1}{4}m^2g_{\tau i} C^\alpha C_\alpha-\frac{1}{2} m^2 C_{\tau}C_{i}+ c_2 H_{\tau i}\\
\end{split}
\end{equation}
\begin{equation}\label{eq:ctaui}
\begin{split}
H^{(0)}_{\tau i}&=-\frac{1}{2}\cancel{ g_{\tau  i}}\,\nabla_\beta C^\beta \nabla_\alpha C^\alpha +(\nabla_\tau C_ i+\nabla_ i C_\tau) \nabla_b C^b+\Big[-\nabla_\tau\left(C_ i\nabla_b C^b\right)-\nabla_ i\left(C_\tau \nabla_\beta C^\beta\right)+\cancel{g_{ i\tau}}\nabla^\sigma\left(\nabla_\alpha C^\alpha C_\sigma\right)\Big]\\
&=(\nabla_\tau C_ i+\nabla_ i C_\tau) \nabla_\beta C^\beta+\Big[-\nabla_\tau\left(\nabla_\beta C^\beta C_ i\right)-\nabla_ i\left(C_\tau\nabla_\beta C^\beta\right)\Big]\\
&=(\partial_i {C_\tau}-2\Gamma^\tau_{i \tau} {C_\tau})\nabla_a C^a+ \Gamma^\tau_{\tau i} {C_\tau} \nabla_b C^b-\Big(\partial_i ({C_\tau} \nabla_b C^b)   - \Gamma^\tau_{i\tau}{C_\tau} \nabla_b C^b )\Big)\\
&=-{C_\tau}[\partial_i(\nabla\cdot C)]
\end{split}
\end{equation}

\subsection{Computational details for case II}\label{app:non-proca3}
\begin{equation}
\mathcal{L}_{(0)}
= R-\frac{1}{4}F_{\mu\nu}\chi^{\mu\nu}-\frac{1}{2}m^2 C_\mu C^\mu+ c^\prime (\nabla_\mu C^\mu)^2 -d^\prime R_{\mu\nu} C^{\mu}C^\nu .
\end{equation}
Equation of motion w.r.t. the metric $g_{\mu\nu}$
\begin{equation}
 \begin{split}
E_{\mu\nu}^{(0)}&=R_{\mu\nu}-\frac{1}{2}g_{\mu\nu}R+\frac{1}{8}g_{\mu\nu} \chi_{\alpha \beta} \chi^{\alpha \beta}-\frac{1}{2}g^{\alpha\beta}\chi_{\mu\alpha}\chi_{\nu\beta}+\frac{1}{4}m^2g_{\mu\nu} C^\alpha C_\alpha-\frac{1}{2} m^2 C_{\mu}C_{\nu}+c' H_{\mu\nu}-d^\prime D_{\mu\nu}
 \end{split}
\end{equation}
where,
 \begin{equation}
\begin{split}
H^{(0)}_{\mu\nu}&=-\frac{1}{2} g_{\mu \nu}\,\nabla_\alpha C^\alpha \nabla_\beta C^\beta +(\nabla_\mu C_\nu+\nabla_\nu C_\mu) \nabla_\alpha C^\alpha-\nabla_\mu\left(C_\nu\nabla_\alpha C^\alpha\right)-\nabla_\nu\left(C_\mu\nabla_\beta C^\beta\right)\\&+g_{\nu\mu}\nabla^\sigma\left(C_\sigma\nabla_\beta C^\beta \right)
    \end{split}
\end{equation}
and,
\begin{equation}\label{eq:Dmunu}
\begin{split}
D^{(0)}_{\mu\nu}=\frac{1}{2} \Big[ 
& - C^{\alpha} C^{\beta} g_{\mu \nu} R_{\alpha \beta}
- \left( \nabla_{\mu} C_{\nu} + \nabla_{\nu} C_{\mu} \right) \left( \nabla_{\alpha} C^{\alpha} \right)
- C^{\alpha} \left( \nabla_{\alpha} \nabla_{\mu} C_{\nu} \right)
- C^{\alpha} \left( \nabla_{\alpha} \nabla_{\nu} C_{\mu} \right) \\
& + C_{\nu} \left( 2 C^{\alpha} R_{\mu \alpha}
- \nabla_{\alpha} \nabla_{\mu} C^{\alpha} + \nabla^{\alpha} \nabla_{\alpha} C_{\mu} \right)
+ C_{\mu} \left( 2 C^{\alpha} R_{\nu \alpha}
- \nabla_{\alpha} \nabla_{\nu} C^{\alpha} + \nabla^{\alpha} \nabla_{\alpha} C_{\nu} \right) \\
& + C^{\alpha} g_{\mu \nu} \left( \nabla_{\alpha} \nabla_{\beta} C^{\beta} \right)
- \left( \nabla_{\nu} C_{\alpha} - 2 \nabla_{\alpha} C_{\nu} \right) \left( \nabla^{\alpha} C_{\mu} \right)
- \left( \nabla_{\mu} C_{\alpha} \right) \left( \nabla^{\alpha} C_{\nu} \right) \\
& + g_{\mu\nu} \left( \nabla_{\alpha} C^{\alpha} \right) \left( \nabla_{\beta} C^{\beta} \right)
+ C^{\alpha} g_{\mu\nu} \left( \nabla_{\beta} \nabla_{\alpha} C^{\beta} \right)
+ g_{\mu\nu} \left( \nabla_{\alpha} C_{\beta} \right) \left( \nabla^{\beta} C^{\alpha} \right)
\Big]
\end{split}
\end{equation}
Evaluating $E_{\tau i}$ component, we get
\begin{equation}
\begin{split}
E_{\tau i}^{(0)}&=R_{\tau i}-\frac{1}{2}g_{\tau i}R+\frac{1}{8}g_{\tau i} \chi_{\alpha \beta} \chi^{\alpha \beta}-\frac{1}{2}g^{\alpha\beta}\chi_{\tau\alpha}\chi_{i\beta}+\frac{1}{4}m^2g_{\tau i} C^\alpha C_\alpha-\frac{1}{2} m^2 C_{\tau}C_{i}+c^\prime C_{\tau i}-d^\prime D_{\tau i}\\
&=R_{\tau i}-\frac{1}{2}g^{\alpha\beta}\chi_{\tau\alpha}\chi_{i\beta}-\frac{1}{2} m^2 C_{\tau}C_{i}\,+ c^\prime C_{\tau i}-d^\prime D_{\tau i}\\
&=-\frac{\partial_iX^{(0)}}{2}-\frac{1}{2}g^{\alpha\beta}\chi_{\tau\alpha}\chi_{i\beta}-\frac{1}{2} m^2 C_{\tau}C_{ i}\,+ c^\prime C_{\tau i}-d^\prime D_{\tau i}\,.
\end{split}
\end{equation}
So, for the validity of the zeroth law, all the terms except the first must vanish for some constraint on the vector field $C_\mu$.
 From \eqref{eq:ctaui}
\begin{equation}
    \begin{split}
     H^{(0)}_{\tau i}&=-{C_\tau}[\partial_i(\nabla.C))]
    \end{split}
\end{equation}
\begin{equation}\label{eq:Dtaui}
\begin{split}
D^{(0)}_{\tau i}& =\frac{1}{2}\bigg[C_{\tau} \Big( 
   2 C^{\alpha} \, R_{i\alpha} 
   + \nabla_{\alpha}\nabla^{\alpha} C_{i} 
   - \nabla_{\alpha}\nabla_{i} C^{\alpha}
\Big) - C^{\alpha}\, \nabla_{\alpha}\nabla_{i} C_{\tau} 
+ C_{i} \Big( 
   2 C^{\alpha}\, R_{\tau \alpha} 
   + \nabla_{\alpha}\nabla^{\alpha} C_{\tau} 
   - \nabla_{\alpha}\nabla_{\tau} C^{\alpha}
\Big) \\
& - C^{\alpha}\, \nabla_{\alpha}\nabla_{\tau} C_{i} 
- (\nabla^{\alpha} C_{\tau}) \Big( -2 \nabla_{\alpha} C_{i} + \nabla_{i} C_{\alpha}\Big) - (\nabla^{\alpha} C_{i})(\nabla_{\tau} C_{\alpha})
- (\nabla_{\alpha} C^{\alpha})(\nabla_{i} C_{\tau} + \nabla_{\tau} C_{i}) \,\bigg]
\\&=\frac{1}{2} \Big[{ C_{\tau}} \bigg( 2 C^{\alpha} R_{i \alpha}
- \nabla_{\alpha} \nabla_{ i} C^{\alpha} + \nabla^{\alpha} \nabla_{\alpha} C_{ i} \bigg) + C_{ i} \bigg( 2 (C_\tau R_{\tau \rho}+C^j R_{\tau j})+
\partial_j (-\nabla_\tau C^j+ \nabla^j C_\tau)\\&+ \Gamma^\tau_{j \tau} (\nabla_\tau C^j-\nabla^j C_\tau) \bigg)- { C_{\tau}} \left( \nabla_{\rho} \nabla_{ i} C_{\tau} +  \nabla_{\rho} \nabla_{ \tau} C_{i}  \right)
- C_{\rho} \left( -\Gamma^\tau_{\tau \tau} \nabla_{\tau} C_{ i}-\Gamma^\tau_{\tau i} \nabla_{\tau} C_{ \tau}-\Gamma^\tau_{\tau \tau} \nabla_{i} C_{ \tau}-\Gamma^\tau_{\tau i} \nabla_{\tau} C_{ \tau}  \right)\\&- C^{j} \bigg(\partial_j (\nabla_\tau C_i)-\Gamma^\tau_{j \tau} (\nabla_\tau C_i)-\Gamma^\tau_{j i}(\nabla_\tau C_\tau)+\partial_j (\nabla_i C_\tau)-\Gamma^\tau_{j \tau} (\nabla_i C_\tau) \bigg)
\\&
- \bigg( \nabla_{ i} C_{\tau}\nabla^{\tau} C_{\tau}+ \nabla_{ i} C_{\rho}\nabla^{\rho} C_{\tau}+ \nabla_{ i} C_{j}\nabla^{j} C_{\tau} - 2 (\nabla_{\tau} C_{ i}\nabla^{\tau} C_{\tau}+\nabla_{\rho} C_{ i}\nabla_{\tau} C_{\tau}+\nabla_{j} C_{ i}\nabla^{j} C_{\tau}) \bigg)
\\&- \bigg(\left( \nabla_{\tau} C_{\tau} \right) \left( \nabla^{\tau} C_{ i} \right)+\left( \nabla_{\tau} C_{\rho} \right) \left( \nabla_{\tau} C_{ i} \right)+\left( \nabla_{\tau} C_{j} \right) \left( \nabla^{j} C_{ i} \right)\bigg)- \left( \nabla_{\tau} C_{i} + \nabla_{ i} C_{\tau}\right) \left( \nabla_{\alpha} C^{\alpha} \right)\Big]\,.
\end{split}
\end{equation}
%%
%%%
Now, if we look at all the terms of $D^{(0)}_{\tau i}$, all the terms except $C_i C^j R_{\tau j}$ have one of the below listed terms:
\begin{itemize}
\item $C_\tau$\,,
\item $\nabla_\tau C_i = \cancelto{0}{\partial_\tau C_i} - \Gamma^\tau_{i \tau} C_\tau$\,,
\item $\nabla_i C_\tau = \partial_i C_\tau - \Gamma^\tau_{i \tau} C_\tau$\,,
\item $\nabla_\tau C_\tau = \cancelto{0}{\partial_\tau C_\tau} - \Gamma^\tau_{\tau \tau} C_\tau$\,.
\end{itemize}
Therefore, $D_{\tau i}$ can be written schematically as a combination of three types of terms,

\begin{equation}
    D^{(0)}_{\tau i}= 2 C_i C^j R^{(0)}_{\tau j} + P C^{(0)}_\tau + Q \partial_i C^{(0)}_\tau
\end{equation}
where \(P\) and \(Q\) denote, respectively, the coefficients of \(C_\tau\) and \(\partial_i C_\tau\), whose explicit forms are algebraically involved. Thefeore, total $E_{\tau i}$ becomes
\begin{equation}
\begin{split}
E^{(0)}_{\tau i}&=-\frac{\partial_iX^{(0)}}{2}-\frac{1}{2}g^{\alpha\beta}\chi^{(0)}_{\tau\alpha}\chi^{(0)}_{i\beta}-\frac{1}{2} m^2 {C^{(0)}_\tau}C^{(0)}_{ i}\,-c^\prime {C^{(0)}_\tau}[\partial_i(\nabla.C)]^{(0)}-d^\prime  (2 C_i C^j R^{(0)}_{\tau j} + P C^{(0)}_\tau + Q \partial_i C^{(0)}_\tau )\\
&=-\frac{\partial_iX^{(0)}}{2}-\frac{1}{2}g^{\alpha\beta}\chi^{(0)}_{\tau\alpha}\chi^{(0)}_{i\beta}-\frac{1}{2} m^2 {C^{(0)}_\tau}C^{(0)}_{ i}\,-c^\prime {C^{(0)}_\tau}[\partial_i(\nabla.C)]^{(0)}-d^\prime  (-\partial_j X^{(0)} C^{(0)}_i C_{(0)}^j + P \,C^{(0)}_\tau + Q\, \partial_i C^{(0)}_\tau )\,.
\end{split}
\end{equation}
The equation of motion w.r.t. the vector field becomes
\begin{equation}
E^{(0)}_\mu =\nabla^\nu \chi_{\nu\mu}-m^2 C_\mu\,-2c^{\prime} \nabla_\mu (\nabla_\nu C^\nu)-2d^\prime R_{\mu\nu} C^\nu\,.
\end{equation}
If we evaluate the $\tau$ component of the above equation, we obtain,
\begin{equation}\label{eq:etaucase1}
 \begin{split}
E^{(0)}_\tau&=\nabla^\nu \chi^{(0)}_{\nu\tau}-b_1m^2 C^{(0)}_\tau\,-\cancel{2c^\prime\nabla_\tau(\nabla_\nu C^\nu)}-2d^\prime R^{(0)}_{\tau\nu}C_{(0)}^\nu.\\
&=h^{ij}_{(0)}\mathcal{D}^{(0)}_i \chi^{(0)}_{j\tau}-b_1 m^2 C^{(0)}_\tau-2 d^\prime \bigg[C^{(0)}_\tau( -\partial_\rho X^{(0)}- \tfrac{1}{2} \omega^{(0)}_i \omega_{(0)}^i
\\&+ \tfrac{1}{2} \partial_i \omega^{i}_{(0)}
- \tfrac{1}{4}\partial_\rho h_{ij} h^{ij} X^{(0)}
+ \tfrac{1}{2} \Gamma^j_{jk}\,\omega^{k}_{(0)} )-\frac{1}{2}h^{ij}\partial_i X^{(0)} C^{(0)}_j\bigg]\\
&=h^{ij}_{(0)}\mathcal{D}^{(0)}_i \chi^{(0)}_{j\tau}-b_1m^2 C^{(0)}_\tau-2 d^\prime \bigg [C^{(0)}_\tau \mathcal{M}-\frac{1}{2}h^{ij}\partial_i X^{(0)} C^{(0)}_j\bigg]
 \end{split}
\end{equation}
where we have defined
\begin{equation}
\mathcal{M}
= -\partial_\rho X^{(0)}- \tfrac{1}{2} \omega^{(0)}_i \omega_{(0)}^i
+ \tfrac{1}{2} \partial_i \omega^{i}_{(0)}
- \tfrac{1}{4}\partial_\rho h_{ij} h^{ij} X^{(0)}
+ \tfrac{1}{2} \Gamma^j_{jk}\,\omega^{k}_{(0)} 
\end{equation}
and, we have computed $R_{\tau\nu}C^\nu$ using \eqref{eq:Rtaui} and \eqref{eq:Rtaurho}
\begin{equation}
\begin{split}
R_{\tau\nu}C^\nu&=\cancel{R_{\tau\tau}}C^\nu+R_{\tau\rho}C^\rho+R_{\tau i}C^i\\
&=\left(-\partial_\rho X-\frac{1}{2} \omega^i \omega_i -\frac{1}{4} \partial_\rho h_{ij} h^{ij} X +\frac{1}{2} \partial_i \omega^i +\frac{1}{2} \Gamma^j_{jk} \omega^k\right) C_\tau-\frac{1}{2}\partial_i X C^i\,.
\end{split}
\end{equation}

\bibliographystyle{JHEP}
\bibliography{Proca}

\end{document}